\documentclass[12pt]{article}
\pdfoutput=1

\usepackage{epsfig}
\usepackage{graphicx}
\usepackage{epstopdf}
\usepackage{amsfonts}
\usepackage{amssymb}
\usepackage{amsthm}
\usepackage{amsmath}
\usepackage{multirow}
\numberwithin{equation}{section} 
\newcommand{\newc}{\newcommand}
\newc{\ra}{\rightarrow}
\newc{\lra}{\leftrightarrow}
\newc{\be}{\begin{equation}}
\newc{\ee}{\end{equation}}
\newc{\bg}{\begin{gathered}}
\newc{\eg}{\end{gathered}}
\newc{\bs}{\begin{split}}
\newc{\es}{\end{split}}
\newc{\ba}{\begin{eqnarray}}
\newc{\ea}{\end{eqnarray}}
\newc{\ov}{\overline}
\newc{\pa}{\partial}
\newc{\D}{\Delta}

\newc{\nn}{\nonumber}

\newc{\tref}[1]{Table \ref{#1}}
\newc{\eref}[1]{Equation \eqref{#1}}
\newc{\fref}[1]{Figure \ref{#1}}
\newc{\sref}[1]{Section \ref{#1}}
\newc{\su}[1]{$SU(#1)$}
\newc{\bm}[1]{\mathbf{#1}}

\begin{document}
\begin{titlepage}

\vspace*{0.7cm}

\begin{center}
{
\bf\LARGE
 Phenomenological implications of a minimal F-theory GUT with discrete symmetry }
\\[12mm]
Athanasios~Karozas$^{\dagger}$
\footnote{E-mail:\texttt{akarozas@cc.uoi.gr}},
Stephen~F.~King$^{\star}$
\footnote{E-mail: \texttt{king@soton.ac.uk}},
George~K.~Leontaris$^{\dagger}$
\footnote{E-mail: \texttt{leonta@uoi.gr}},
Andrew~K.~Meadowcroft$^{\star}$
\footnote{E-mail: \texttt{a.meadowcroft@soton.ac.uk}}
\\[-2mm]

\end{center}
\vspace*{0.50cm}
\centerline{$^{\star}$ \it
School of Physics and Astronomy, University of Southampton,}
\centerline{\it
SO17 1BJ Southampton, United Kingdom }
\vspace*{0.2cm}
\centerline{$^{\dagger}$ \it
Physics Department, Theory Division, Ioannina University,}
\centerline{\it
GR-45110 Ioannina, Greece}
\vspace*{1.20cm}

\begin{abstract}
\noindent
We discuss the origin of both non-Abelian discrete family symmetry and Abelian
continuous family symmetry, as well as matter parity, from F-theory SUSY GUTs.
We propose a minimal model based on the smallest GUT group
$SU(5)$, together with the non-Abelian family symmetry $D_4$ plus an Abelian
family symmetry, where fluxes are responsible for doublet-triplet splitting, leading to a realistic
low energy spectrum with phenomenologically acceptable quark and lepton masses and mixing.
We show how a $Z_2$ matter parity emerging from F-theory can suppress proton decay
while allowing neutron-antineutron oscillations, providing a distinctive signature of the set-up.
 \end{abstract}

 \end{titlepage}

\thispagestyle{empty}
\vfill
\newpage

\setcounter{page}{1}

\section{Introduction }
F-theory~\cite{Vafa:1996xn} models have attracted considerable interest over the recent years~\cite{Beasley:2008dc}-\cite{Grimm:2009yu}.
For example, Supersymmetric (SUSY) Grand Unified Theories (GUTs) based on $SU(5)$
has been shown to emerge naturally from F-theory~\cite{Dudas:2009hu}-\cite{Cvetic:2015txa}.
However, in the F-theory context, the $SU(5)$ GUT group
is only one part of a larger symmetry. The other parts
manifest themselves at low energies as Abelian and/or non-Abelian discrete symmetries,
which can be identified as family symmetries,
leading to significant constraints in the effective superpotential.

In this paper we review the basic mechanisms responsible for the origin
of both non-Abelian discrete family symmetry and Abelian
continuous family symmetry, as well as matter parity, from F-theory SUSY GUTs,
before piecing together the first realistic example model of its kind which includes
all three types of symmetries.
In order to make this paper self-contained, and hopefully useful to model builders not familiar
with F-theory, we shall include necessary introductory material, as well as a discussion of
basic features which will be obvious to F-theory experts, but may be new to the expected readership
of this paper. Thus we begin with a fairly general introduction (or a reminder for the experts) on
how symmetries of any kind can emerge from F-theory (for more details see
 reviews~\cite{Denef:2008wq}-\cite{Leontaris:2015yva}).

\begin{figure}\centering
\includegraphics[scale=0.6]{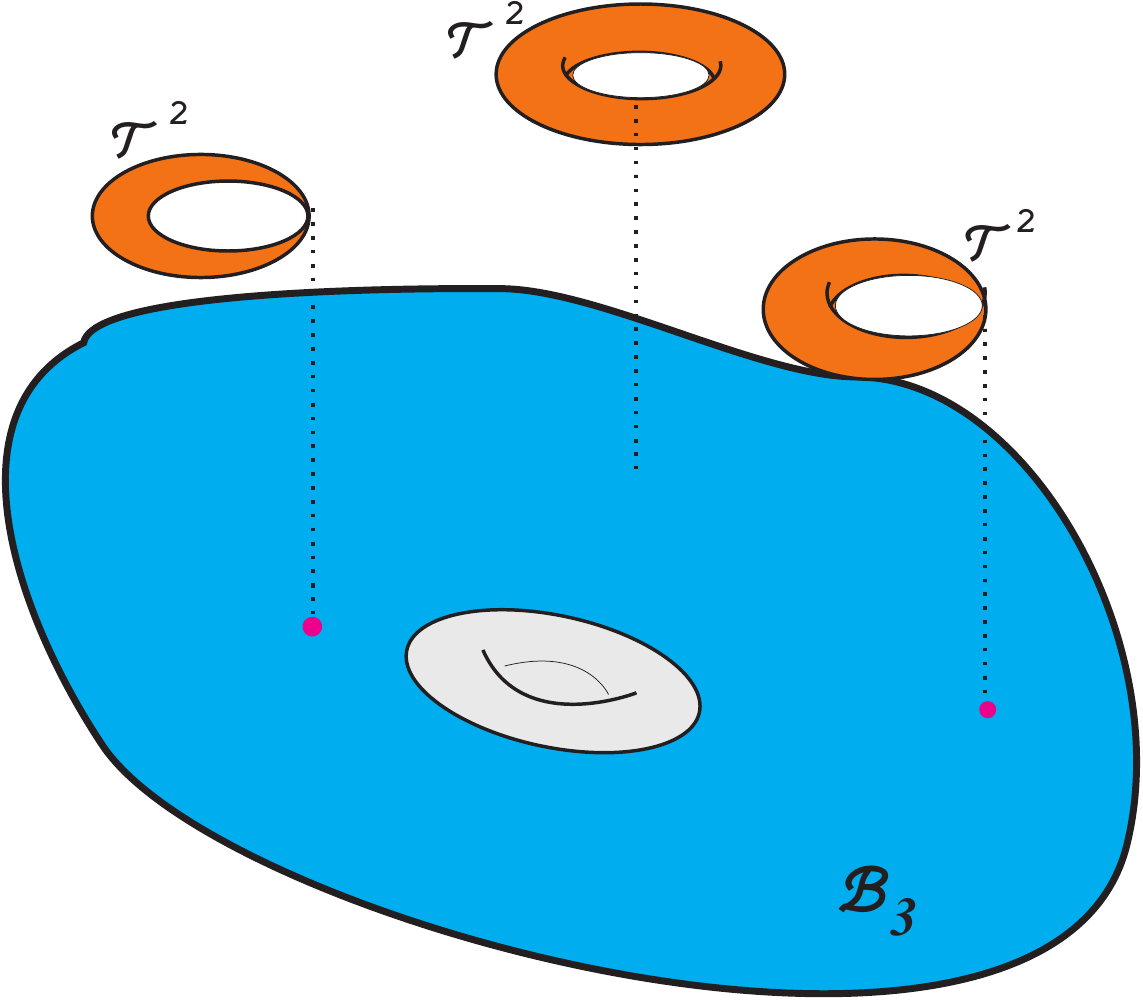}
\caption{A pictorial representation of a Calabi fourfold, which exhibits elliptic fibration over a threefold base, $B_3$. The fibration is manifest as a 2-Torus at every point in the base, as shown. The modulus of the torus at each point is related to the axio-dilaton profile, $\tau=C_0+i/g_s$. Where the fibre degenerates, the presence of a D7-brane orthogonal to the base is indicated. \label{cy4f}}
\end{figure}

As a basic starting point, it is worth remarking that
F-theory is a non-perturbative formulation of type IIB string theory
invariant under a $SL(2,Z)$ symmetry (the $S$-duality) which attains a
geometric realisation. Current F-theory constructions are based on an elliptically
fibred internal space where the complex  modulus of the elliptic fiber
is a combination of
the axion and dilaton fields $\tau = C_0+i e^{-\phi}$, i.e., the two
scalars of the type IIB bosonic spectrum (see \fref{cy4f}).
This way, F-theory can be considered as a 12-dimensional string theory
compactified on a torus characterised by the above modulus $\tau$.
Algebraically, the fibration is described by a birationally equivalent
complex cubic equation,  the so called Weierstra\ss\, model \cite{Beasley:2008dc,Donagi:2008ca,Beasley:2008kw}.
Depending on the specific structure of its coefficients,
at certain points of the fibration the torus degenerates and the
fibration becomes singular.
All possible singularities have been classified with respect to the
vanishing order of
the coefficients (polynomials) and the discriminant of the
Weierstra\ss\, equation.
  It was shown long time ago~\cite{Kodaira} (see also recent works~\cite{Bershadsky:1996nh,Esole:2011sm})
 that these singularities are of ADE type (in the Cartan classification of
non-Abelian groups),
the highest being the  ${ E}_{8}$ exceptional group.

In F-theory non-Abelian gauge symmetries are linked to the
singularities of the elliptic fibre.
Hence, old successful  GUTs based on the exceptional groups ${ E}_{6,7,8}$,
as well as the lower rank $SO(10)$ and $SU(5)$ ones,
can be naturally realised as effective F-theory models.  As such, they
constitute  a particularly promising component of the vast string landscape,
since many  parameters
of the effective low energy models are determined from a few basic
topological
properties of the compact space associated to the geometric nature of the
singularities.

However, one might object that building a model from F-theory one has
to deal with
complications due to the as yet unknown global geometrical structure of
the internal space. Moreover,  various mathematical  issues of the
elliptic fibrations,
whose role in model building is not well understood, would further
obscure physics.
Despite the complicated structure of the global geometry, it is often
adequate to  focus on
a local F-theory  description where computations  are simpler leading
to reliable predictions
of the effective model's  parameter space.

In the local approach, one may
associate the GUT symmetry
to a particular divisor of the elliptically fibred manifold and use
techniques such as
the spectral cover \cite{Donagi:2008ca} to deal with the implications of the remaining symmetry and
the topological properties of the compact space.
In this context,  one may determine the massless spectrum of the
effective theory
and all its properties under the GUT group and its quantum numbers
with respect to the symmetries of the spectral cover.
  Furthermore from general characteristics of the compact manifold and
G-fluxes~\cite{Donagi:2008ca}
  we can  determine the chiralities of the massless spectrum.

Within the local approach, we can focus on a small patch and compute several
  important quantities such as the Yukawa couplings~\cite{Heckman:2008qa}-\cite{Marchesano:2015dfa}. Indeed, in F-theory
massless fields reside on the intersections of various D7-branes,
usually called
matter curves.  In this picture, a massless state is described by a
wavefunction which exhibits a
Gaussian profile picked on the corresponding matter curve and can be
determined by solving
the appropriate equations of motion.
The Yukawa couplings   occur at triple intersections of three matter curves.
Studying locally  the  wavefunctions' profiles  of the relevant states
one is  able  to compute the strength of these couplings and predict the mass
spectrum and (in principle)
  all possible  interactions allowed within a specific model.

  In addition to the non-Abelian sector, in F-theory effective models
are endowed
  with Abelian and discrete symmetries which may  arise   either as a
subgroup of the
  non-Abelian symmetry or from a non-trivial Mordell-Weil
  group associated to rational sections of the elliptic fibration.
  It is well known that the discrete symmetries in particular
are extremely important in  suppressing undesired proton decay operators
  and generate a hierarchical fermion mass spectrum~\footnote{For discussions in a wider framework of
  discrete symmetries in String Theory see references~\cite{BerasaluceGonzalez:2011wy}-\cite{Cvetic:2015moa}.}.
Furthermore, non-Abelian discrete groups were introduced to
interpret the mixing  properties of
the neutrino sector~\cite{Altarelli:2010gt,Ishimori:2010au,King:2013eh,King:2014nza}.

In the present work, then, we will focus on non-Abelian discrete symmetries emerging
in the context of the spectral cover, accompanied by continuous Abelian symmetry.
We continue to investigate the grid of discrete symmetries emerging as
subgroups of the $SU(5)_{\perp}$ spectral cover symmetry.
Motivated by the successful implementation of a class of such symmetries to
the neutrino sector, we focus on the subgroups of $S_4$.
We also show how a geometric discrete $Z_2$ symmetry can additionally emerge,
leading to matter parity which can control proton decay operators.
However, due to the basic feature of F-theory constructions with flux breaking
of the GUT group yielding doublet-triplet splitting and
incomplete GUT representations, the matter parity is necessarily of a new kind.
In the particular example we develop, based on
$D_4\times U(1)$ family symmetry, with an $SU(5)$ GUT group, broken by fluxes,
the geometric $Z_2$ matter parity, while suppressing proton decay,
allows neutron-antineutron oscillations, providing a distinctive signature of the set-up.
To be precise,  while $QL d^c$ is forbidden, the operator $u^cd^cd^c$ is present
leading to $n\bar n$ oscillations at a calculable rate.

The layout of the remainder of the paper is as follows.
In Section~\ref{basics}, we review the basics of F-theory GUTs based on $SU(5)$
together with additonal $U(1)^4$ groups, or subgroups thereof.
In Section~\ref{spectral} we describe the Spectral Cover approach and show how
a $S_4\times U(1)$ group can emerge from this formalism; we also discuss the origin of
an additional $Z_2$ geometric symmetry which can play the role of matter parity. In Section~\ref{extrasec}
 we clarify in some detail the action of the possible monodromy group on the matter representations, 
 distinguishing  abelian and non-abelian discrete cases.
In Section~\ref{D4} we show how a $D_4$ discrete symmetry  subgroup of the $S_4$ can emerge.
The structure of this non-Abelian discrete symmetry seems promising, so can be used to illustrate
in the simplest setting many of the features of interest, and can be used as the basis for constructing
a realistic model which we do in Section~\ref{model}.
In Section~\ref{baryon} we investigate the physics of baryon number violation in this model,
showing how the combination of symmetries can suppress proton decay, but allows
baryon number violating operators which can yield neutron-antineutron oscillations, providing
a distinctive signature of our scheme.
Section~\ref{conclusions} concludes the paper.

\section{F-$SU(5)$ basics with $U(1)^4$}
\label{basics}
In this work we are interested in $SU(5)_{GUT}$ which is the minimum
GUT group accommodating the Standard Model symmetry.
This is embedded in a local $E_8$ singularity according to, such that the massless spectrum is found in adjoint of $E_8$, which  under
the maximal decomposition of  the $E_8\to  SU(5)_{GUT} \times SU(5)_{\perp} $
decomposes as follows
\ba
  248 & \rightarrow & (24,1)+(1,24)
+(10,5)+(\overline{5},10)+(\overline{10},
\overline{5})+(5,\overline{10}).
  \label{E8SU5}
\ea
As expected,  the $ SU(5)_{GUT}$  multiplets have transformation
properties under the second $SU(5)_{\perp} $.
The $10$ multiplets in particular are in the fundamental of
$SU(5)_{\perp} $, while the
$5$-plets are in the antisymmetric  of $SU(5)_{\perp}$. Depending on
the geometry of the
internal manifold and the fluxes, $SU(5)_{\perp}$ can be broken to an
appropriate subgroup.
This way, matter curves and hence,
the $SU(5)_{GUT}$ representations acquire specific topological and
symmetry properties inherited
to the fermion families and Higgs fields.

There are a variety of symmetry options embedded in  $SU(5)_{\perp}$
and our choice
in this work will be dictated by observational facts.
As already discussed, we will focus on non-Abelian discrete symmetries
accompanying
$SU(5)_{GUT}$. Nevertheless, to set the stage, it is convenient first
to start with the Abelian symmetries.
In this case we have the following breaking pattern
\be
E_8  \supset SU(5)\times SU(5)_\perp \rightarrow SU(5)\times U(1)_\perp^{4}.
\label{DP0}
\ee
  Thus, the $SU(5)_{GUT}$ matter content transforms non-trivially
under the Cartan subalgebra
of $SU(5)_{\perp}$ with weight vectors $t_{1,...,5}$ satisfying
\be
\sum_it_i=0.\label{SU5trace}
\ee
   Under the above  notation the matter curves accommodating the
   $SU(5)$  representations are labelled as follows
   \be
   10_{t_i}, \;   \overline{10}_{-t_i}, \;   \bar 5_{t_i+t_j},\;
5_{-t_i-t_j},\;  1_{t_i-t_j}
   \ee
   where, due to antisymmetry, the indices of the fiveplets must differ $i\ne j$.
    As a result, the `charges' $t_i$ distinguish the various
   matter curves and eventually the fermion generations associated to some of them.

In principle  the superpotential can be constrained by all these four
$U(1)$'s, however monodromy 
actions reduce their number while the constraints are adjusted
accordingly. In general, monodromies are
necessary in order to allow a diagonal tree-level coupling for the
top-quark. Indeed, any  $SU(5)_{GUT}$
invariant  trilinear coupling should respect the $U(1)$ symmetries. In
this case, the following tree-level
couplings can be realised
\ba
10_{t_i} 10_{t_j} 5_{-t_i-t_j},\; 10_{t_i}\bar 5_{t_j+t_k}\bar
5_{t_l+t_m},\; 10_{t_i}\overline{10}_{-t_j}1_{t_j-t_i}.
\ea
The first  term contributes to the up-quark sector, while the $U(1)$
invariance is guaranteed by the fact that  the `charges'  sum up to zero:
 $\{t_i\}+\{t_j\}+\{-t_i-t_j\}=0$.   The second term is $U(1)$ invariant
due to (\ref{SU5trace}) provided
all indices $i,j,k,l,m$ are different. The third term might prove
useful to provide heavy masses to extra tenplet  pairs.

A few remarks are in order. Firstly, if all $U(1)$ symmetries are unbroken,
the coupling $10_{t_i} 10_{t_j} 5_{-t_i-t_j}$ contributes only to non-diagonal
mass terms, thus there is no diagonal  top-quark coupling as required. The reason is that
due to antisymmetry we must have $i\ne j $. Secondly,  it is not
possible to generate a term
coupling additional fiveplet-antifiveplet pairs, since this would
require singlets with charges
${t_k+t_l-t_i-t_j}$.
\ba
\bar 5_{t_i+t_j} 5_{-t_k-t_l}1_{t_k+t_l-t_i-t_j}\label{1outE8}
\ea
  Such singlets might exist only outside the $E_8$ whose heterotic
duals   might be associated  with non perturbative states~\cite{Baume:2015wia}.

However, as already mentioned, there is an action on $t_i$'s of a
non-trivial monodromy group.
The  minimal possibility is a  ${ Z}_2$ monodromy,
$t_1\leftrightarrow t_2$, i.e, the one which
identifies two $U(1)$ charges. This leads to an identification of the
corresponding matter curves,
where the tenplets reside. As a result, the coupling
\[ 10_{t_1} 10_{t_2} 5_{-t_1-t_2} \to 10_{t_1} 10_{t_1} 5_{-2t_1} \]
becomes diagonal and a top-quark mass is obtained from a tree-level coupling.

Moreover, certain couplings of the type $5 \cdot \bar 5\cdot 1$ which
within the original
symmetry structure would require a singlet of the type given in
(\ref{1outE8}),
after the monodromy action are in principle allowed with singlets
within $E_8$ matter.
  Indeed, for a $Z_2$ monodromy such that $t_j=t_k $ for example,
(\ref{1outE8})
becomes
\[ \bar 5_{t_i+t_j} 5_{-t_j-t_l}1_{t_l-t_i}\]
Therefore, in contrast to the singlet field of the term
(\ref{1outE8}), here the same fiveplet pair
receives a mass with a singlet vev $1_{t_l-t_i}$ which is embedded in
$E_8$.  We observe that,  monodromies are capable of generating
important Yukawa terms (like $\mu$-term), making -at
least in some cases- the additional singlets redundant.

 As already noted, the reduction of the accompanying symmetry
to the maximum number of abelian factors $SU(5)_{\perp}\to U(1)^4$  discussed above is
just one of a plethora of possibilities.   There exists a variety of 
groups embedded in $SU(5)_{\perp}$ which can in principle appear in the
effective theory. These include non-Abelian groups $SU(n)$ with  $n<5$, 
or discrete ones, as well as combinations of both of them. 

As for the corresponding geometrical picture, 
in  the spectral cover approach, the Higgs bundle over the
GUT divisor is described
by a five degree polynomial, $\sum_kb_ks^{5-k}=0$, whose roots are
associated to the $t_i$ charges and
its coefficients carry the information from the background geometry.
These roots may fall into a
  variety of monodromy groups and as long as the discrete options are
concerned, this can be
a Galois subgroup of the Weyl group $W(SU(5)_{\perp})=S_5$. This
includes  the alternating
groups ${ A}_n $, the dihedral groups ${ D}_n$ and cyclic
groups ${ Z}_n$, $n\le 5$ and the
Klein four-group ${ Z}_2\times { Z}_2$.  
We will discuss these options in the next sections.

\section{The Spectral Cover approach}
\label{spectral}

Before going into our phenomenological investigation of the available list of discrete symmetries accompanying the $SU(5)$ GUT,   
in this section we recapitulate a few general  aspects of the spectral cover approach. 

F-theory is characterised as a Calabi-Yau complex fourfold, elliptically fibred over a three complex dimensional space, $B_3$ with three complex coordinates $x,y,z$, corresponding to the six compact spatial dimensions.
The GUT gauge group lives on the del Pezzo surface with coordinates
$x,y$.
The elliptic fibration is described mathematically by the Weierstra\ss\, equation,
\be y^2=x^3+f(z)x+g(z)\,,\label{weierstrass}\ee
where $f(z)$ and $g(z)$ are eighth and twelfth degree polynomials in $z$, the coordinate normal to the GUT surface. The fibre of this space has singularities that can be classified by examining the zeroes of the discriminant and the vanishing order of the functions $f(z), g(z)$ of equation \eqref{weierstrass}. These singularities have been classified by Kodaira~\cite{Kodaira}
(see also ~\cite{Bershadsky:1996nh,Esole:2011sm}),  and are in general associated with non-Abelian gauge groups, with the largest allowed symmetry group being the exceptional group $E_8$.
In the current work we will be assuming an $SU(5)_{GUT}$, which has a corresponding commutant with the $E_8$ 
of $SU(5)_{\perp}$, called the perpendicular group. It is within this perpendicular $SU(5)_{\perp}$
group that the extra symmetries, commuting with the $SU(5)_{GUT}$, reside.

\subsection{Spectral Cover of the 10s}
The spectral cover equation for the $10$s of the \su{5} GUT group are defined by a fifth order polynomial.
In this scenario, the Weierstra\ss\, equation can be recast into the so-called Spectral Cover equation~\cite{Donagi:2009ra}:
\be C_5:\,\sum_k^5b_ks^{5-k}=0 \,,\ee
where $b_k$ are holomorphic sections and $s$ is an affine parameter derived  from the coordinates on the underlying manifold. From Group theory we know that $SU(5)_{\perp}$ can be decomposed into four $U(1)$ factors. As such we can suppose that the spectral cover be allowed to factorise into a product of irreducible parts. For example, we will take an interest in the case where one of the roots factorises out to a linear part:
\be \label{fact10} C_5\rightarrow C_4\times C_1:\,(a_5s^4+a_4s^3+a_3s^2+a_2s+a_1)(a_6+a_7s)\,.\ee
The remaining four roots, are considered to be related under some non-trivial monodromy group action. The most general of these would be $S_4$, so we shall assume this to start with. Later we will examine how to obtain a subgroup of $S_4$. Notice that, due to the assumed factorisation, there is also an accompanying
$U(1)$ continuous Abelian group associated with the fifth root in the $C_1$ factor.

For the most general, $S_4$, monodromy action we need only take the polynomial equation describing the tenplets of the $SU(5)_{GUT}$ group, Equation (\ref{fact10}), with no further constraints. The zeroth order coefficient adequately describes the matter curves available in this scenario, as the zeros of $b_5\sim\prod_it_i$ determine the localisation of the matter multiplets. This trivially gives us two matter curves as $b_5=a_1a_6=0$, which is in keeping with a monodromy relating four of the roots of the polynomial. We should also make use of the tracelessness condition of \su{5},
\be b_1=a_5a_6+a_4a_7=0, \ee
which is equivalent to the sum of the roots equaling zero. This can be solved by the introduction of a parameter $a_0$, such that:
\be\label{tsol} \left\{a_4\to a_0 a_6,a_5\to -a_0 a_7\right\}\ee
It can be shown that this does not introduce extra sections to the spectral cover equation.
\subsection{Spectral Cover of the 5s}
The spectral cover equation for the $5$s of the \su{5} GUT group are defined by a tenth order polynomial in a similar way to the $10$s,
\be C_{10}:\,\sum_{n=1}^{10}c_ns^{10-n}=b_0\prod_{i<j}(s-t_i-t_j)\,.\ee
Since the roots of this equation are related to those of the $C_5$ spectral cover equation, we can express the defining equation (the zeroth order coefficient $c_{10}$) in terms of the coefficients of $C_5$. This gives the standard equation for the $5$s~\cite{Donagi:2008ca}:
\be \label{fives} P_5=b_4b_3^2-b_2b_5b_3+b_0b_5^2\,.\ee

\noindent Expressing the $b_{k}$ in terms of $a_{j}$  the equation of the fiveplets factorises as

\begin{equation}
\label{5fvs}P_5=\overbrace{\left(a_2^2a_7+a_2a_3a_6- a_0a_1a_6^2\right)}^{P_a}\underbrace{\left(a_3 a_6^2+(a_2a_6+a_1a_7)a_7\right)}_{P_b}\,.
\end{equation}

\noindent In order to obtain this result we also used the relation \eqref{tsol}.

\subsection{A note on monodromies and Yukawas}

We have already pointed out that a monodromy action is required to achieve a tree-level Yukawa coupling
supporting a heavy top-quark.
 But when monodromies are introduced, the theory also becomes more complicated.
In the context of  the eight-dimensional theory there exists an adjoint Higgs paremetrising
the ``normal'' direction  to the GUT divisor. In the simplest case, we can take the Higgs vev
to take values  along  the Cartan subalgebra. But when monodromy actions are assumed,
then this simple description with Higgs vev values along the diagonal, is inadequate.
The generalisation is to take Higgs backgrounds which are no-longer diagonalisable,
but they generally assume  a block-diagonal form:
\[  {\cal H}= {\cal H}_1\oplus {\cal H}_2\oplus \cdots \oplus{\cal H}_k\]
where each Higgs component $ {\cal H}_j$ is represented by a $n_j\times n_j$ matrix.
There is a corresponding splitting of the monodromy group $G=\prod_j G_j$ where each
component acts via Weyl permutations on the corresponding block of the Higgs field~\cite{Cecotti:2009zf}.
Correspondingly, the spectral cover ${\cal C}$  would factorise in an equivalent
number of factors, $\prod_j {\cal C}_j $ where every spectral surface ${\cal C}_j $  
is associated to a corresponding polynomial factor.  We note in passing that
the  notion of monodromy is very useful in the computation of the top Yukawa coupling,
however, we will not discuss this issue in the present work. In the subsequence, we will restrict our analysis to the case where the physics of the effective model is captured by the spectral cover polynomial ${\cal C}_5$. Depending on specific conditions the polynomial factorises to a number of factors $\prod_j{\cal C}_j$, where each spectral surface ${\cal C}_j$ is associated to the Weyl group acting on the corresponding polynomial roots. We  describe the corresponding picture in the next sections.

\subsection{Matter Parity from Geometric Symmetries\label{gps}}
In addition to symmetries contained within $SU(5)_{\perp}$,
the spectral cover equation admits a geometric symmetry that will impose constraints on the coefficients of the equation. This symmetry will be of the $Z_N$ type, which may serve as a matter parity, preventing unwanted dimension four proton decay operators. In this subsection, we give a brief introduction to this geometric symmetry, for more details see Appendix D,  as has been discussed in~\cite{Hayashi:2009bt}.

Given that, up to a phase, the spectral cover equation is invariant under:
\begin{align}\sigma:\,\,\,\,\,\,\,\,\,\, s&\rightarrow s \text{e}^{i\phi}\\ \,\,\,\,b_k&\rightarrow b_k \text{e}^{i(\chi+(k-6)\phi)}\end{align}
we may enforce this symmetry on the $a_i$ coefficients also. This can be achieved by  extending  the symmetry to the line bundles associated to the matter and Higgs representations of $SU(5)$.\\

 The defining equation of each  matter curve is written in terms of coefficients which arise from suitable factorisations of the coefficients $b_k$. For our particular factorisation $ C_4\times C_1$ we start from the relations $b_k(a_{m})$,
which for our present purposes can be written collectively as follows
\be  b_k \propto a_{n} a_{m}\label{banda}\ee
with the indices subject to the constraint $ k+n+m =12 $. Let's assume that under the above mapping  $a_{n}$ transforms as:
\be a_{n} \to  \text{e}^{i\psi_{n}} \text{e}^{i (3-n)\phi} a_{n} \ee
so that the product  $a_{n} a_{m}$ picks up a total phase:
\be a_{n} a_{m} \to  \text{e}^{i(\psi_{n}+\psi_{m})} \text{e}^{i (6-n-m)\phi} a_{n}a_m= \text{e}^{i(\psi_{n}+\psi_{m})} \text{e}^{-i (6-k)\phi} a_{n}a_m\ee
where we have recalled that $n+m=12-k$. This shows that this is a consistent implementation of this symmetry for the $a_k$ coefficients, provided:
\be \chi= \psi_{n}+\psi_{m}\ee
This is easily done since the two phases $\psi_{n},\psi_{m}$ are independent of the index $k$ and can be adjusted accordingly. However we should note that not all $a_k$ are created equal, meaning that our $\psi$ are not entirely independent. For example, by consistency, we must have the following:
\be\bg b_5=a_1a_6\\b_4=a_2a_6+a_1a_7\eg\ee
From this we can see that since each $b_k$ must have a general phase of $\chi$, independent of $k$, the products $a_1a_6$, $a_2a_6$ and $a_1a_7$ must have phases summing to $\chi$. Clearly, it is necessary that $\chi=\psi_1+\psi_6$ by the first condition, but it is also required that $\chi=\psi_2+\psi_6$. Thus, $\psi_1=\psi_2$ and similarly as $\chi=\psi_1+\psi_7$, $\psi_6=\psi_7$. We may extend this argument such that we have only two allowed phases: $\psi_i=\psi_1=\psi_2=\dots=\psi_5$ and $\psi_j=\psi_6=\psi_7$.\\

Having obtained the transformation properties of $a_{n}$, we can determine now those of the matter curves using their defining equations in a trivial requirement of consistency.

\section{Discrete symmetries from the spectral cover}
\label{extrasec}

In order to facilitate the analysis in the next sections,  here we will summarise previous work on the issue
of abelian and non-abelian discrete symmetries. Our presentation will rely mainly on the
work of ref~\cite{Heckman:2009mn} as well as in \cite{Marsano:2009gv} and  especially
 \cite{Antoniadis:2013joa} where non-Abelian fluxes are conjectured to give rise to 
 non-Abelian discrete family symmetries in the low energy effective theory. 
 The origin of such a symmetry is the non-Abelian $SU(5)_{\perp}$
which accompanies $SU(5)_{GUT}$ at the $E_8$ point of enhancement. Whether a non-Abelian
symmetry survives in the low energy theory will depend on the geometry of the compactified space and the
fluxes present. The usual assumption is that the $SU(5)_{\perp}$ is first broken to 
a product of $U(1)_{\perp}$ groups which are then further broken by the action of discrete symmetries associated with the monodromy group. Instead here we are following the conjecture in  
\cite{Antoniadis:2013joa} that  {\em non-Abelian fluxes} can break $SU(5)_{\perp}$ first to a non-Abelian discrete group $S_4$ then to a smaller group such as $D_4$ which acts as a family symmetry group in the low energy effective theory. We emphasise that this is a conjecture since there is no proof that non-Abelian fluxes can do this. In these  works discrete symmetries were used to deal with fundamental problems of
 the effective model such as the neutrino sector, the $\mu$ term etc. 
 
  In F-theory compactifications there exist discrete group actions corresponding to certain monodromies around
 the singularities.  
 In such a case  we obtain  an effective field theory model where various matter curves 
of the original theory are related by  the action of a  finite group associated to the singularity. 
Depending on the specific geometric symmetry of the compactification,  the monodromy group could be
a $Z_n$ group or a non-Abelian discrete symmetry such as the permutation symmetries $S_n, A_n, D_n$.
The action of a $Z_n$ group will map the matter fields to one another, while in the non-Abelian
discrete cases there are non-trivial representations accommodating  the matter fields of the same orbit.
There are various phenomenological reason suggestive for a non-Abelian discrete  symmetry.
In the context of F-theory in particular,  the $D_4$ symmetry
was suggested in~\cite{Heckman:2009mn}  for a successful implementation of a consistent effective model.
This was considered in the context of a model where all Yukawa hierarchies emerge from a single $E_8$ enhancement point.
It was further shown that the $D_4$  symmetry is one of the few possible monodromy
groups accommodating just only the minimal matter, and at the same time being compatible with viable
right-handed neutrino scenarios.
In the present work, we will try to exploit the non-abelian nature of this discrete group in order to
construct viable fermion mass textures which interpret the neutrino data  and make possible predictions
for other interesting processes of our effective model.

In our general discussion of the previous  section we have seen that the $SU(5)$ multiplets accommodating the SM
spectrum are distinguished under the charges $t_i$ associated to the four $U(1)$ factors embedded into
 $SU(5)_{\perp}$. The requirement for rank-one up-quark mass matrix is met by appealing to an exchange symmetry,
 the minimal being the one which identifies two tenplets $10_{t_a}\leftrightarrow 10_{t_b}$.
  This is equivalent to a $Z_2$ symmetry which maps one matter curve to the other.   In fact, such symmetries are generic since
 seven branes are found at the singularities of the fiber where the symmetry is enhanced. The associated geometrical
 structure is  described by polynomial equations and the relevant  properties are encoded in the coefficients of the latter
 \cite{Heckman:2009mn}. Hence, depending on the specific features of the geometry, these coefficients may exhibit properties
 such that the polynomial  roots are related to non-trivial symmetries beyond the  $Z_2$  described above.

Since our gauge field theory model is based on the $SU(5)_{GUT}$, any exchange symmetry must appear in
 the context of the $SU(5)_{\perp}$, which arises  in the case of the maximal decomposition of the $E_8$ singularity.
 Any discrete symmetry expected in the above context, must be a subgroup of the maximal  Weyl group
under $SU(5)$, which is $S_5$ -  the group of all permutations on a set of five elements.
Now, we recall that the $SU(5)_{GUT}$ representations reside along matter curves that are
characterised by the elements in the Cartan of $SU(5)_{\perp}$. According to our previous analysis
these elements are just the five roots $t_{1,2,3,4,5}$ of the corresponding spectral cover
polynomial. In effect, additional properties of each $SU(5)_{GUT}$ matter curve
are attributed to the particular exchange symmetries of these weights.
If we assume the most generic geometry, then  $S_5$ would represent the monodromy group
leading to rather restrictive  identifications of the matter curves.
Because of phenomenological constraints, we should consider less restrictive geometries
relying on some suitable  subgroup of $S_5$.  This would imply specific relations
or  identifications among a fraction of the original matter curves, leaving the remaining
curves intact.

We can approach the above picture from the point of view of the spectral
polynomial    ${\cal C}_5: \; \sum_k b_k s^{5-k}=0  $ whose roots are the weights $t_i$.
We know that the properties of  $b_k$ coefficients are well defined by the geometry.
In order to specify the properties of the roots $t_i$
we note that  $b_k$ are symmetric functions of roots $b_k=b_k(t_i)$,  however the
solutions $t_i=t_i(b_k)$ are in general non-trivial and may imply the existence of
a monodromy group which is identified as the Galois group of the roots.
For any  monodromy group which is smaller that $S_5$ there is a corresponding factorisation
of the spectral cover polynomial.  The possible ways of factorising the spectral polynomial ${\cal C}_5$  are:
\[    {\cal C}_4\times {\cal C}_1,\; \;      {\cal C}_3\times {\cal C}_2,\; \;
 {\cal C}_3\times {\cal C}_1^2,\; \;     {\cal C}_2\times {\cal C}_2\times{\cal C}_1,\;\;{\cal C}_2\times{\cal C}_1^3\,.\]
The above factorisations imply non-trivial constraints on the superpotential of the
effective model.
  \begin{figure}[!t]
  \centering
 {\flushleft \includegraphics[scale=0.4]{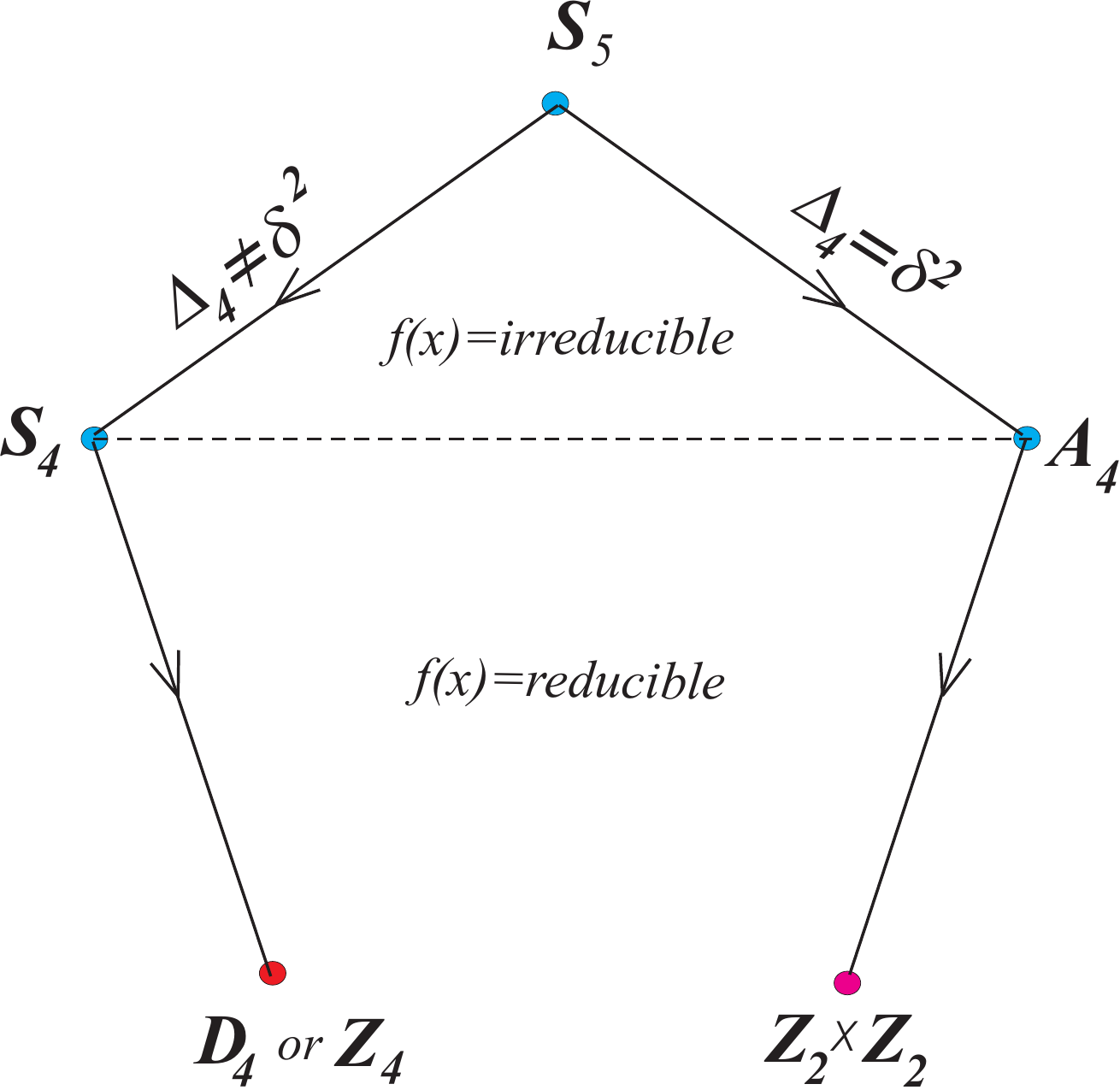}}
  \caption{A pictorial form of the reduction of the discrete group $S_5 $ to
   subgroups $S_4, A_4$ with respect   to the properties of the corresponding
    $\Delta_4$ Discriminant and the resolvent.  See text for details.}
  \label{GalGroups}
  \end{figure}
As already pointed out in the introduction, guided from phenomenological reasons
in this  work we will analyse the $ {\cal C}_4\times {\cal C}_1$ case.  For this
case the splitting of the five-degree polynomial is given in equation  \eref{fact10}
where the coefficients $a_i$ of the new polynomials are related  to
$b_k$ in a straightforward manner. We have already explained how these relations
 determine  the homologies  of the new coefficients from those of $b_k$'s and
 discussed their implications on the effective theory in the previous sections.
 However, there are additional interesting features of these coefficients
 with respect to the monodromy groups which we now analyse.
 For our case of interest, the non-trivial part is related to the
 fourth order polynomial so that  the maximal symmetry group $S_5$ reduces down to
 $S_4$, (i.e. the permutation of four objects), or to some of its subgroups.

  The precise determination of the Galois group depends solely on the
  specific structure of the coefficients $a_n, n=1,\dots, 5$.
  Leaving the details for the appendix, we only state here that
  they can be classified in terms of symmetric functions of the roots.
  Concerning the particular symmetry groups we are dealing with, it suffices to examine  the Discriminant $\Delta_4$
  and the resolvent of the corresponding fourth-degree polynomial.

The discriminant $\Delta_4$ is a symmetric function of the roots $t_{1,2,3,4}$ and as such
it can always be expressed as a function of the coefficients $a_i$, hence $\Delta_4=\Delta_4(a_i)$.
For generic coefficients $a_i$ the symmetry is $S_4$ unless
  $\Delta_4$ can be written as a square of a quantify $\delta(a_i)$ which
  is invariant under the $S_4$-even permutations which constitute the group $A_4$.

The resolvent is the cubic polynomial
\be 
 f(x)\,=\, a_5^{\frac{3}{2}} (x-x_1)(x-x_2) (x-x_3)\sim x^3+c_1 x^2+c_2x+c_3 \label{cubicres}
 \ee
where the roots $x_{1,2,3}$ are  the three $t_i$-combinations
\begin{equation}
x_{1}=t_{1}t_{2}+t_{3}t_{4},\quad{x_{2}=t_{1}t_{3}+t_{2}t_{4}},\quad{x_{3}=t_{3}t_{2}+t_{1}t_{4}} .
\label{rootss}
\end{equation}
These are invariant under the group $D_4$ which is the symmetry of the square.
It can be seen that all coefficients  $c_k$ of the polynomial are symmetric
functions of $t_i$ and therefore they can also be expressed as functions of $a_i$,
$c_k(a_i)$.
Depending on the specific properties of the two quantities $\Delta_4$ and $f(x)$, the Galois group may be
any of the $S_5$ subgroups depicted in figure~\ref{GalGroups}.

We can readily deduce
that -depending on the reducibility of the resolvent-
the Galois group of the roots is either a non-Abelian or Abelian discrete group.

From the point of view of the low energy effective theory, there is a clear distinction
between the two categories of discrete groups.  As is well known,
non-abelian discrete groups are endowed with non-trivial (non-singlet) representations.
In effect, ordinary GUT representations transform non-trivially under such symmetries.
This way, additional restrictions might be imposed on superpotential terms
while specific forms of mass textures may arise at the same time.
In the subsequent, we will focus on the particular discrete symmetry of $D_4$.

\section{The discrete group $D_4$ as a Family Symmetry}
\label{D4}

In a realistic F-theory effective  model a  superpotential should emerge containing all
 necessary interaction terms. In particular it should   distinguish the three families and provide
 correct  masses to  all fermion fields and at the same time should  exclude all other undesired ones.  
In the previous sections, it has become evident that the imperative distinction of
the three fermion  families,  in F-theory should be related to possible
 additional  symmetries and geometric properties carried by the $SU(5)$ matter curves. 
In this section we will continue to explore the origin of such symmetries 
in the context of the ${\cal C}_4\times {\cal C}_1$ splitting. 
With respect to the corresponding gauge group,  
we may turn on suitable Abelian and non-Abelian fluxes which result in the
 breaking of the  $SU(5)_{\perp}$ symmetry.  Hence, 
in the present case for example one can turn on a flux along a non-trivial line bundle of 
the corresponding Cartan $U(1)$ so that the group originally breaks to $SU(4)_{\perp}\times U(1)_{\perp}$.
Furthermore,  one may assume  the reduction of the  $SU(4)_{\perp}$ part 
to some  discrete group, as a consequence of a suitable non-Abelian flux or appropriate Higgsing.
The case of $D_4\subset S_4$ in particular  can be reached from our initial maximal symmetry of ${\cal C}_4$
under the following chain $ SU(4)\to SU(3) \to S_4$. 
Indeed, we may invoke the one-to-one correspondence~\cite{Luhn:2007yr} of the  $S_4$ representations
to  those of $SU(3)$, and decompose the $SU(4)$ ones   according to the pattern shown in
Table~\ref{43S4}.
   \begin{table}[t] \centering%
\begin{tabular}{ccccl}
$SU(4)$& $\subset$& $SU(3)$&$\subset$&$S_4$\\
$4$&$\to$& $3+1$&$\to$&$3+1$\\ 
$6$&$\to$&$3+\bar 3$&$\to$&$3+3$\\ 
$15$&$\to$&$8+1+3+\bar 3$&$\to$&$3+3'+2+1$\\
\end{tabular}%
\caption{The embedding of $S_4$ representations in  the  ${\cal C}_4$ spectral cover symmetry
} \label{43S4}
\end{table}

An analogous symmetry reduction could be attained through the Higgs bundle description and 
in particular the spectral cover of the fundamental and anti-symmetric representations
of our GUT gauge group. In this local picture we may exploit the fact that the geometric
singularities  essentially correspond to particular symmetries of the effective theory model.  
Hence, in accordance with the choice of the family group in our previous discussion, we 
will appeal to local geometry and assume the non-abelian discrete  group 
$D_4$ acting on the $SU(5)_{GUT}$ representations. 
 To study its implications in our particular construction we start by splitting the five roots $t_i$  into  
 two sets
 \[{\cal C}_4\leftrightarrow \{t_{1}, t_2, t_3,  t_4\},\;\;  \;\; {\cal C}_1\leftrightarrow \{t_5\}  \]
in accordance with our choice of  spectral cover splitting. Because $SU(5)_{GUT}$ representations are characterised
by the weights $t_i$, as a result they fall into  appropriate orbits. Hence, the matter curves accommodating
the tenplets   of $SU(5)_{GUT}$ are
\ba
10_{a}&:& \{t_{1}, t_2, t_3,  t_4\}\nonumber\\
10_{b}&:& \{t_{5}\}\nonumber
\ea
In the same way, if no other restrictions are imposed, the  matter curves for
the fiveplets   of $SU(5)$ also fall into two categories
\ba
\bar 5_{c}&:& \{t_1+t_2, t_2+t_3, t_3+t_4, t_4+t_1, t_1+t_3, t_2+t_4 \}\nonumber\\
\bar 5_{d}&:& \{t_1+t_{5}, t_2+t_{5}, t_3+t_{5}, t_4+t_{5}\}.\nonumber
\ea
We can readily observe that the orbits are `closed' under the action of the elements of the $S_4$ group.
The $SU(5)$ superpotential couplings are subject to constraints in accordance with the above
classification. Hence, for example the $10_a 10_b 5_d$ coupling is allowed  while   $10_a 10_b 5_c$ is incompatible
with the $S_4$ rules.

 The invariance of the orbits under the action of the whole set of $S_4$  elements reflects
the fact that the polynomial coefficients $a_k$ of the corresponding spectral cover fourth-order polynomial
are quite generic. On the contrary, if specific restrictions are imposed on $a_k$ the discrete group
will be a subgroup of $S_4$, while further splitting of the orbits will occur.  We will now be more 
specific and consider the case  of the dihedral group, $D_4\subset S_4$.

In the context of F-theory with an $SU(5)$ GUT group, if the left-over discrete group is  $D_4$, then the 
four of the roots of the original $SU(5)_{\perp}$
 group are permuted in accordance to the specific $D_4$  rules  and the overall symmetry structure is:
\begin{align*} E_8\rightarrow& SU(5)_{\text{GUT}}\times SU(5)_{\perp}\\
\rightarrow& SU(5)_{\text{GUT}}\times D_4\times U(1)_{\perp}\,.
\end{align*}\\

In order to have a $D_4$ symmetry relating the four roots, rather than an $S_4$, we must appeal to Galois theory.
From \tref{Galoisgroup} in the Appendix A, we can see that this means the discriminant of the quartic part of \eref{fact10} must not be a square,
while the cubic resolvent of the polynomial must be reducible.

If we assume the roots $t_{i=1,2,3,4}$, then the quartic part of \eref{fact10} has a cubic resolvent of the  form
given in~(\ref{cubicres})
where the roots $x_i$ are the symmetric polynomials of the weights  $t_i$   given in~(\ref{rootss}).

It can be shown that the discriminant ($\Delta_f$) of \eref{cubicres} is:
\be 27 \Delta_f=4 \left(a_3^2-3 a_2 a_4+12 a_1 a_5\right){}^3-\left(2 a_3^3-9 \left(a_2 a_4+8 a_1 a_5\right) a_3+27 \left(a_5 a_2^2+a_1
   a_4^2\right)\right){}^2\ee
which is also equal to the discriminant of the quartic polynomial relating the four roots - this is a standard property of all cubic resolvents\footnote{An alternative cubic resolvent  
is presented in the Appendix C.}. \\

By computing the coefficients as functions of the $a_{i}$ coefficients, the cubic takes the form
\begin{equation}
f(s)=a_{5}^{-\frac{3}{2}}[(a_{5}s)^{3}-a_{3}(a_{5}s)^{2}+(a_{2}a_{4}-4a_{1}a_{5})a_{5}s+(4a_{1}a_{3}a_{5}-a_{2}^{2}a_{5}-a_{1}a_{4}^{2})]\,.
\end{equation}

\noindent The simplest way to make this polynomial reducible, is to demand the zero order term to vanish, $f(0)=0$. This means that one of the roots equals to zero. 
 By setting $f(0)=0$ and using the \su{5} tracelessness constraint ($b_{1}=0$\footnote{Note that $b_1=a_5a_6+a_4a_7=0$ is solved as shown in \eref{tsol}}) we take the following known condition~\cite{Antoniadis:2013joa} between the $a_{i}$'s :

\begin{equation}
a_{2}^{2}a_{7}=a_{1}(a_{0}a_{6}^{2}+4a_{3}a_{7})\,,
\end{equation}

\noindent If we then substitute this into the equation for the fiveplets of the GUT group, \eqref{5fvs}, we get an equation factorised into 3 parts,

\begin{equation}
P_{5}=a_{3}(a_{2}a_{6}+4a_{1}a_{7})(a_{3}a_{6}^{2}+a_{7}(a_{2}a_{6}+a_{1}a_{7}))\,,
\end{equation}
which indicates that we have at least 3 distinct matter curves by the usual interpretation.
\begin{table}[t!]\centering
\begin{tabular}{|c|c|c|}\hline
$SU(5)$ Rep.&Equation&Homology\\\hline
$10_a$&$a_1$&$\eta -5c_1 -\chi$\\
$10_b$&$a_6$&$\chi$\\
$5_a$&$a_3$&$\eta -3c_1 -\chi$\\
$5_b$&$a_2a_6+4a_1a_7$&$\eta-4c_1$\\
$5_c$&$a_3a_6^2+a_7(a_2a_6+a_1a_7)$&$\eta -3c_1 +\chi$\\\hline
\end{tabular}

\caption{Summary of the default matter curve splitting from spectral cover equation in the event of a $D_4$ symmetry
	 accompanying an \su{5} GUT group in the case of the symmetric polynomials $x_{i=1,2,3}$ as discussed in text.\label{unsplit}}
\end{table} 

\noindent The so obtained splittings of the non-trivial $SU(5)$ representations are collected in Table 2. The first column indicates
the $SU(5)$ representation, while the  defining equation of each
corresponding matter curve is shown in column 2. In the third column we designate the associated homologies. These are readily determined
from the known Chern classes
 of the $b_k$ coefficients through the equations $b_k=b_k(a_i)$ given in (\ref{banda}), using well known
procedures~\cite{Dudas:2010zb,King:2010mq}. These are expressed in terms of the known classes~\footnote{The Chern class of $b_k$ is $[b_k]= (6-k) c_1-t=\eta -kc_1$ where $c_1$ is the first Chern class of the GUT ``divisor'' $S$ and $-t$ the corresponding one of the normal bundle~\cite{Donagi:2009ra}.} $\eta, c_1$ and an arbitrary one designated by $\chi$.

\subsection{Irreducible Representations}
Thus far we have largely ignored how the group theory must be applied to matter curves in this construction. We shall now examine this side of the problem, with a particular view being taken to find the irreducible representations where possible. Given the earlier conjecture that  {\em non-Abelian fluxes} can break $SU(5)_{\perp}$ to $D_4$, which acts as a family symmetry group in the low energy effective theory, it then follows that the low energy states must transform according to irreducible representations of $D_4$.
In Appendix A we show how reducible 4 and 6 dimensional representations
of $D_4$ decompose into irreducible representations. The argument in Appendix A is summarised as follows.


Knowing that we have four weights $t_{i=1,2,3,4}$, that have a relation under a $D_4$ symmetry, we might exploit the  nature of $D_4$. Specifically, since $D_4$ can be physically interpreted as a square, we might label the corner of such a square with our weights and see how they must transform based on this. It is then clear that there should be two generators for this symmetry: a rotation about the centre by $\frac{\pi}{2}$ and a reflection along one of the lines of symmetry, which we will call $a$ and $b$ respectively. This is in keeping with the presentation of $D_4$:
\be a^4=e,\,\,\,b^2=e,\,\,\,bab=a^{-1}\,,\ee
where $e$ is the identity.\\

It can be shown that this quadruplet of weights can be rotated into a basis with irreducible representations of $D_4$ - see Appendix A - by use of appropriate unitary transformations. It transpires that the irreducible basis includes a trivial singlet, a non-trivial singlet and a doublet, as summarised in \tref{Irreps10}. Note that we also have an extra singlet that is charged under the fifth weight ($10_\delta$), which must logically be a trivial singlet since it is uncharged under the $D_4$ symmetry.\\

\begin{table}[t!]\centering
\begin{tabular}{|l|c|c|}\hline
Curve&$D_4$ rep.&$t_5$\\\hline
$10_{\alpha}$&$1_{++}$&0\\
$10_{\beta}$&$1_{+-}$&0\\
$10_{\gamma}$&$2$&0\\
$10_{\delta}$&$1_{++}$&1\\
\hline
\end{tabular}\caption{Table summarising the representations of the tens of $SU(5)_{\text{GUT}}$ \label{Irreps10}}
\end{table}

The $\bar 5/5$ representations of the GUT group have a maximum of 10 weights   before the reduction of the $SU(4)\to D_4$ symmetry. These have weights related to the $10$s of the GUT group: $\pm (t_i+t_j)$. By consistency these must transform in the same manner as the weights of the $10$s, allowing us to unambiguously write down the generators $a$ and $b$. \\

By the same process as before, we may decompose this tenplet under $D_4$ into irreducible representations of the group. Referring to the Appendix once again, we may obtain a total of eight representations, as shown in \tref{Irreps5}. However, we note that three of the representations\footnote{$\bar{5}_{\delta}$, $\bar{5}_{\epsilon}$, and $\bar{5}_{\zeta}$} are entirely indistinguishable as they are trivial singlets with only charges under $t_{i=1,2,3,4}$.\\

A full decomposition of the $SU(5)_{GUT}$ representations is included in Appendix A, including the block diagonalisation procedure as applied to the singlets of the group, which will be important for model building in this work. 

\begin{table}[t!]\centering

\begin{tabular}{|c|c|c|c|}\hline
Curve&$D_4$ rep.&$t_5$ charge&weight relation\\\hline
$\bar{5}_{\alpha}$&$1_{++}$&1&$\sum_{i=1}^4 t_i$\\
$\bar{5}_{\beta}$&$1_{+-}$&1&$(t_1+t_3)-(t_2+t_4)$\\
$\bar{5}_{\gamma}$&$2$&1&$\left(\begin{array}{c}t_1-t_3\\t_2-t_4\end{array}\right)$\\\hline
$\bar{5}_{\delta}$&$1_{++}$&0&$\sum_{i=1}^4 t_i$\\
$\bar{5}_{\epsilon}$&$1_{++}$&0&$\sum_{i=1}^4 t_i$\\
$\bar{5}_{\zeta}$&$1_{++}$&0&$\sum_{i=1}^4 t_i$\\
$\bar{5}_{\eta}$&$1_{+-}$&0&$(t_1+t_3)-(t_2+t_4)$\\
$\bar{5}_{\theta}$&$2$&0&$\left(\begin{array}{c}t_1-t_3\\t_2-t_4\end{array}\right)$\\
\hline
\end{tabular}\caption{Table summarising the representations of the fives of $SU(5)_{\text{GUT}}$ \label{Irreps5}}
\end{table}

\subsection{Reconciling Interpretations}
It is clear at this point that there is some tension between the two angles of attack for this problem. Obviously we must be able to describe both the non-abelian discrete group representations of the matter curves, while also being able to obtain them in some manner from the spectral cover approach. In order to achieve this, we shall attempt some form of multifurcation of the spectral cover by definition of new sections in a consistent manner. \\

Let us begin by defining two new sections $\kappa,\lambda$ such that
   \be a_3\to\kappa a_7,\;\; a_2\to\lambda\,a_6\,. \ee
   It is clear that this approach has some similarity with the tracelessness constraint solution usually employed (\eref{tsol}). Furthermore, these definitions do not  generate new unwanted sections. For example, the $b_k$'s
   \be b_0=-a_0 a_7^2, b_1=0, b_2= a_7^2 \kappa +a_0 a_6^2, b_3= (\kappa +\lambda )a_6 a_7, b_4=\lambda a_6^2  +a_1 a_7, b_5=a_1 a_6\,,
   \ee
   do not acquire an overall common factor,
    while the discriminant
   \be\Delta =108 a_0 \left(\lambda a_6^2  +4 a_1 a_7\right) \left(\kappa ^2 a_7^2 +a_0 \left( \lambda a_6^2 +4
      a_1 a_7\right)\right){}^2\ne \delta^2\ee
     is not a square - as required for the case of a $D_4$ monodromy group. On the contrary, substitution to equation \eqref{5fvs} gives
   \be P_a= a_6^2 \left( (\kappa +\lambda )\lambda a_7 -a_0 a_1\right) \ee
   and
\be P_b=   a_7 \left((\kappa +\lambda ) a_6^2 +a_1 a_7\right)\,.\ee
This appears to generate extra matter curves by increasing the number of factors available, with the added advantage that we can easily find the homologies of our matter curves and know the flux restraints for each. We can interpret these results as a multifurcation to irreducible representations of the $D_4$ group.

If we further assume $a_1\to\mu a_6$, then
\be P_b=  a_6 a_7 \left(a_6 (\kappa +\lambda )+\mu a_7\right)\,,\ee
and the tens of the GUT group now become:
\be P_{10}\to b_5=\mu\,a_6a_6\,.\ee
So we add extra curves here as well.

\begin{table}[t!]\centering\begin{tabular}{|c|c|c|c|}\hline
Constraints & $P_{a}$ & $P_{b}$ & $P_{10}$\\\hline\hline
$a_{1}={\kappa a_{2}}$ & & &\\
$a_{3}={\lambda a_{7}}$ & $a_2^2 \left(a_7+\lambda\mu a_7-\alpha_{0}\kappa\mu^2 a_2\right)$ &$a_2 a_7\left(\kappa a_7+(\lambda\mu+1)\mu a_2 \right)$ & $\kappa\mu a_{2}^2$\\
$a_{6}={\mu a_{2}}$ & & &\\\hline
\end{tabular}\caption{A viable splitting option of the matter curves, respecting the constraint $\Delta\ne \delta^2$
as required for $D_4$ symmetry.
 \label{splits}}\end{table}

This is not a unique choice of splitting, and in fact we have a number of possible options that would be compatible with the requirement to prevent unwanted overall factors. A second option is the splitting:
\be a_1\to \lambda a_2  , \quad a_3\to \kappa a_7 \,.  \ee
With this choice, the fives are now
\be P_a=  a_2 \left(a_7 \left(a_6 \kappa +a_2\right)-a_0 a_6^2 \lambda \right) \ee
\noindent and
\be P_b=
 a_7 \left(a_6^2 \kappa +a_2
   \left(a_7 \lambda +a_6\right)\right)\,.\ee
The tens now reads $P_{10}=a_{1}a_{6}\rightarrow{\lambda a_{2} a_{6}}$.

In the same way we can find a number of combinations that leads in suitable splits.
In Table~\ref{splits}  we show the most interesting case
\be
a_{1}\to\kappa a_{2},\quad a_{3}\to\lambda a_{7},\quad\text{and}\quad a_{6}\to\mu a_{2}.\label{case5}
\ee
As we can see (\ref{case5}) leads in a maximal factorisation for the fives (six factors) and the tens (four factors). The homologies of the new coefficients are

\be
[\kappa]=-c_1,\quad [\mu]=-[\lambda]=4 c_1+2\chi-\eta .
\ee

\noindent Using the above, we can calculate the homologies of the all new factors of tenplets and fives. Notice that the distribution of the the tens and the fives has be done in a arbit  
This case is of particular interest because we have seen that we have four tens of the GUT group, while we will also have six of the fives provided we interpret the trivial singlets as one representation. This last assumption seems reasonable given that they are otherwise indistinguishable.\\

\subsubsection{Flux Restrictions}
\begin{table}[t!]\centering\begin{tabular}{|c|c|c|}\hline
 \multicolumn{3}{|c|}{$P_{10}=\kappa\mu a_{2}^{2}$} \\
  \hline\hline
 Curve & factor & Homology \\\hline
  $10_{1}$ & $\kappa$ & $-c_1$ \\\hline
   $10_2$ & $a_2$ & $\eta-4c_1-\chi$ \\\hline
    $10_3$ & $a_2$ & $\eta-4c_1-\chi$ \\\hline
     $10_4$ & $\mu$ & $-\eta+4c_1+2\chi$ \\
  \hline
\end{tabular}
\caption{Distribution of the tens according to the new factorisation, $P_{10}=\kappa\mu a_{2}^{2}$.}\label{tens}
\end{table}
\begin{table}[t!]\centering\begin{tabular}{|c|c|c|c|}\hline
 \multicolumn{4}{|c|}{$P_{b}=a_2 a_7\left(\kappa a_7+(\lambda\mu+1)\mu a_2 \right)$} \\
  \hline\hline
 Curve & $t_{5}$ charge & factor & Homology \\\hline
  $\bar{5}_{a}$ & $1$ & $a_2$ & $\eta-4c_1-\chi$ \\\hline
   $\bar{5}_{b}$ & $1$ & $a_7$ & $c_1+\chi$ \\\hline
   $\bar{5}_{c}$ & $1$ & $\kappa a_7+(\lambda\mu+1)\mu a_2$ & $\chi$ \\\hline
    \multicolumn{4}{|c|}{$P_{a}=a_2^2 \left(a_7+\lambda\mu a_7-\alpha_{0}\kappa\mu^2 a_2\right)
$}\\\hline\hline
    Curve & $t_{5}$ charge & factor & Homology \\\hline
  $\bar{5}_{d}$ & $0$ & $a_2$ & $\eta-4c_1-\chi$ \\\hline
   $\bar{5}_{e}$ & $0$ & $a_2$ & $\eta-4c_1-\chi$ \\\hline
   $\bar{5}_{f}$ & $0$ & $a_7+\lambda\mu a_7-\alpha_{0}\kappa\mu^2 a_2$ & $c_1+\chi$ \\\hline
\end{tabular}
\caption{Distribution of the fives into $P_a$ and $P_b$. As we can see $P_{b}$ are related with the $t_5$ charge.}\label{fives}
\end{table}

In order to finally marry the two understandings present in this work, we must appeal to flux restrictions. We summarise 
the homologies of the various matter curves in \tref{tens} and \tref{fives} with this in mind.  Let us assume the usual flux restriction rules.
We denote with ${\cal F}_{Y}$ the $U(1)_Y$ flux which  breaks $SU(5)$ to the Standard Model and at the same time generates chirality to
the fermions.  In order  to avoid a Green-Schwarz mass for the corresponding gauge boson we must require ${\cal F}_{Y}\cdot\eta={\cal F}_{Y}\cdot c_1=0$. 
For the unspecified homology $\chi$ we parametrise the corresponding flux restriction with an arbitrary integer $N={\cal F}_{Y}\cdot\chi$,
hence we have the constraints:
\be
\bg {\cal F}_{Y}\cdot\chi=N,\; {\cal F}_{Y}\cdot c_1={\cal F}_{Y}\cdot\eta=0\,.\eg
\ee
We shall also assume the doublet-triplet splitting mechanism to be powered by this flux. 
Indeed, assuming  $N$ units of hyperflux piercing a given matter curve,  the $5$/$\bar{5}$  split according to:
\begin{equation}
\label{5split}
\begin{gathered}n(3,1)_{-1/3}-n(\bar{3},1)_{+1/3}=M_5\,,\\
 n(1,2)_{+1/2}-n(1,2)_{-1/2}=M_5+N\,.\end{gathered}
 \end{equation}
Thus,  as long as  $N\ne 0$,  for the fives residing on a given  matter curved 
the number of doublets differs from the number of triplets  in the effective theory.
Choosing  $M_5=0$ for a Higgs matter curve the coloured triplet-antitriplet fields
appear only in pairs which under certain conditions~\cite{Beasley:2008dc,Donagi:2008ca}
 form heavy massive states. On the other hand,
the difference of the doublet-antidoublet fields is non-zero and is determined solely from the 
hyperflux integer parameter $N$.  Similarly, on a matter curve accommodating fermion generations,
Equation (\ref{5split}) implies different numbers of lepton doublets and down quarks on this
particular matter curve. As a consequence, the corresponding mass matrices are expected to 
differ.
 
Similarly, the  $10$s decompose under the influence of $N$ hyperflux units to the
  following SM-representations:
\begin{equation}
\label{10split}
\begin{gathered}n(3,2)_{+1/6}-n(\bar{3},2)_{-1/6}=M_{10}\,,\\
 n(\bar{3},1)_{-2/3}-n(3,1)_{+2/3}=M_{10}-N\,,\\
n(1,1)_{+1}-n(1,1)_{-1}=M_{10}+N\,.\end{gathered}
\end{equation}
Hence, as in the case of fives above, the flux effects have analogous implications on the
tenplets. The first line in (\ref{10split}) in particular, generates the 
required up-quark chirality since for $M_{10}\ne 0$ the number of $Q=(3,2)_{1/6}$ differs from   $\bar Q=(\bar 3,2)_{-1/6}$
representations.  Moreover, from the second line it is to be observed  that $N\ne 0$ leads to
further splitting between the  $ Q=( 3,2)_{1/6}$ and  $u^c=(\bar 3,1)_{-2/3}$ multiplicities. 
This fact as we will see provides interesting non-trivial quark mass matrix textures.

\begin{table}\centering
\begin{tabular}{|c|c|c|c|}\hline
GUT rep& Def. Eqn.& Parity:& Matter content\\\hline
$10_1$& $\kappa$& $-$&$M_{1}Q_L+u^c_LM_{1}+e^c_LM_{1}$\\
$10_2$& $a_2$& $a$&$M_{2}Q_L+u^c_L(M_{2}+N)+e^c_L(M_{2}-N)$\\
$10_3$& $a_2$& $a$&$M_{3}Q_L+u^c_L(M_{3}+N)+e^c_L(M_{3}-N)$\\
$10_4$& $\mu$& $\frac{parity(a_6)}{a}$&$M_{4}Q_L+u^c_L(M_{4}-2N)+e^c_L(M_{4}+2N)$\\\hline
$5_a$&$a_2$& $a$&$M_{a}\bar{d}^c_L+(M_{a}-N)\bar{L}$\\
$5_b$&$a_7$& $b$&$M_{d}D_u+(M_{d}+N)H_u$\\
$5_c$&$\kappa a_7$& $-b$&$M_{c}\bar{d}^c_L+(M_{c}+N)\bar{L}$\\
$5_d$&$a_2$& $a$&$M_{b}\bar{D}_d+(M_{b}-N)\bar{H}_d$\\
$5_e$&$a_2$& $a$&$M_{e}\bar{d}^c_L+(M_{e}-N)\bar{L}$\\
$5_f$&$a_7$& $b$&$M_{f}\bar{d}^c_L+(M_{f}+N)\bar{L}$\\\hline
\end{tabular}\caption{The Generalized matter spectrum for the model before marrying $D_4$ representations and the matter curves from the spectral cover.}\label{genmat}
\end{table}

\section{Constructing An $N=1$ Model}
\label{model}

Referring to the aforementioned geometric symmetry discussed at length in the Appendix, we may start out by assigning a $Z_2$ symmetry to our matter curves, \tref{gsym}. We shall demand some doublet-triplet splitting in our model, so we take the liberty of setting $N=1$, motivated by a desire to produce a spectrum free of Higgs colour triplets. \\

The $Z_2$ parity has arbitrary phases connecting the coefficients in two cycles: $a_{1,\dots,5}$ and $a_{6,7}$, which we must choose so that we can best fit the standard matter parity. The generalised parities of the matter curves are presented in \tref{genmat}. If we start with  a handful of basic requirements it becomes quickly apparent how to do this and guides our assignments of the $D_4$ irreducible representations.
\begin{enumerate}
\item We must have a tree-level Top Yukawa coupling and no other tree-level Yukawas
\item We wish to forbid Dimension 4 proton decay - which may be achieved if our Higgs have $+$ parity and our matter $-$ parity
\item We want a spectrum that resembles the MSSM
\end{enumerate}
If we examine \tref{gsym}, we can see that in order to be free from $D_{u,d}$ matter, we should choose the parity option $a=b=+$. The subtlety here is that the $H_u$ and $H_d$ must be on matter curves that have different homologies so that if we set the multiplicity for those curves to zero (preventing the $D_{u,d}$ matter), the flux naturally pushes the $H_u$ to be on a $5$ of the GUT group, while it pushes the $H_d$ to be a $\bar{5}$.
\begin{table}\flushleft\footnotesize
\begin{tabular}{|c|c|c|c|c|c|c|c|c|}\hline
GUT rep& Def. Eqn.& Parity:&$(-,-)$&$(+,-)$&$(-,+)$&$(+,+)$&$N=1$ Matter spectrum\\\hline
$10_1$& $\kappa$& $-$&$-$&$-$&$-$&$-$&$M_{1}Q_L+u^c_LM_{1}+e^c_LM_{1}$\\
$10_2$& $a_2$& $a$&$-$&$+$&$-$&$+$& $M_{2}Q_L+u^c_L(M_{2}+1)+e^c_L(M_{2}-1)$\\
$10_3$& $a_2$& $a$&$-$&$+$&$-$&$+$& $M_{3}Q_L+u^c_L(M_{3}+1)+e^c_L(M_{3}-1)$\\
$10_4$& $\mu$& $\frac{parity(a_6)}{a}$&$-$&$+$&$+$&$-$&$M_{4}Q_L+u^c_L(M_{4}-2)+e^c_L(M_{4}+2)$\\\hline
$5_a$&$a_2$& $a$&$-$&$+$&$-$&$+$&$M_{a}\bar{d}^c_L+(M_{a}-1)\bar{L}$\\
$5_b$&$a_7$& $b$&$-$&$-$&$+$&$+$&$M_{d}D_u+(M_{d}+1)H_u$\\
$5_c$&$\kappa a_7$& $-b$&$+$&$+$&$-$&$-$&$M_{c}\bar{d}^c_L+(M_{c}+1)\bar{L}$\\
$5_d$&$a_2$& $a$&$-$&$+$&$-$&$+$&$M_{b}\bar{D}_d+(M_{b}-1)\bar{H}_d$\\
$5_e$&$a_2$& $a$&$-$&$+$&$-$&$+$&$M_{e}\bar{d}^c_L+(M_{e}-1)\bar{L}$\\
$5_f$&$a_7$& $b$&$-$&$-$&$+$&$+$&$M_{f}\bar{d}^c_L+(M_{f}+1)\bar{L}$\\\hline
\end{tabular}\caption{Parity options are $(a=\pm,b=\pm)$. Any matter curve that has a $D_4$-doublet  must produce doublets - i.e. split twice as fast. $a=parity(a_2)$ and $b=parity(a_7)$, by convention.\label{gsym}}
\end{table}
\\

We now select our multiplicities $M_i$ as follows:
\begin{align*}
M_2=&M_3=M_b=M_d=0\\
M_1=&M_a=-M_f=1\\
M_4=&2\\
M_c=&-4
\end{align*}
This provides us with a spectrum that has only a Top Yukawa at tree-level, the correct number of matter generations, and only  $u^cd^cd^c$ type dimension 4 parity violating operators, which should shield us from the most dangerous proton decay operators. The spectrum is summarized in \tref{modeltable}.
\begin{table}[t!]\centering
\begin{tabular}{|c|c|c|c|c|c|}\hline
GUT rep& Def. Eqn.& Parity& Matter content&$D_4$ rep.&$t_5$ charge\\\hline
$10_1$& $\kappa$& $-$&$Q_L+u^c_L+e^c_L$&  $1_{+-}$ &0 \\
$10_2$& $a_2$& $+$&$u^c_L-e^c_L$&   $1_{++}$ &0\\
$10_3$& $a_2$& $+$&$u^c_L-e^c_L$&  $1_{++}$ &1 \\
$10_4$& $\mu$& $-$&$2Q_L+4e^c_L$& $2$ &0 \\\hline
$5_a$&$a_2$& $+$&$2\bar{d}^c_L$&$2$  &0 \\
$5_b$&$a_7$& $+$&$H_u$& $1_{++}$ &0\\
$5_c$&$\kappa a_7$& $-$&$-4\bar{d}^c_L-3\bar{L}$& $1_{+-}$&0 \\
$5_d$&$a_2$& $+$&$-\bar{H}_d$& $1_{++}$ &$-1$ \\
$5_e$&$a_2$& $+$&$\bar{d^c_L}$& $1_{+-}$ &$-1$\\
$5_f$&$a_7$& $+$&$-2\bar{d}^c_L$&  $2$ &$-1$\\\hline
\end{tabular}
\caption{Full spectrum for an $SU(5)\times D_4\times U(1)_{t_5}$ model from an F-theory construct. Note that the $-t_5$ charge corresponds to the $5$, while any representations that are a $\bar{5}$ will instead have $t_5$.\label{modeltable}}

\begin{tabular}{|c|c|c|c|c|}\hline
Singlet&Parity&$D_4$ rep.&$t_5$ charge&Vacuum Expectation\\\hline
$\theta_{\alpha}$&$+$&$1_{++}$&$-1$&$\langle\theta_{\alpha}\rangle=\alpha$\\
$\theta_{\beta}$&$-$&$1_{+-}$&$-1$&$\langle\theta_{\beta}\rangle=\beta$\\
$\theta_{\gamma}$&$+$&$2$&$-1$&$\langle\theta_{\gamma}\rangle=(\gamma_1,\gamma_2)$\\
$\theta_{a}$&$+$&$2$&$0$&$\langle\theta_{a}\rangle=(a_1,a_2)$\\
$\nu_{r}$&$-$&$1_{+-}$&$0$&$-$\\
$\nu_{R}$&$-$&$2$&$0$&$-$\\\hline

\end{tabular}\caption{Spectrum of the require singlets to construct full Yukawa matrices with the model outlined in \tref{modeltable}.\label{modeltable2}}
\end{table}\newpage


\subsection{Operators}
\begin{table}\centering
\begin{tabular}{|c|c|c|c|}\hline
Low Energy Spectrum&$D_4$ rep& $U(1)_{t_5}$&$Z_2$\\\hline
$Q_3,u^c_3,e^c_3$&$1_{+-}$&$0$&$-$\\
$u^c_2$&$1_{++}$&$1$&$+$\\
$u^c_1$&$1_{++}$&$0$&$+$\\
$Q_{1,2},e^c_{1,2}$&$2$&$0$&$-$\\
$L_i, d^c_i$&$1_{+-}$&$0$&$-$\\
$\nu^c_3$&$1_{+-}$&$0$&$-$\\
$\nu^c_{1,2}$&$2$&$0$&$-$\\
$H_u$&$1_{++}$&$0$&$+$\\
$H_d$&$1_{++}$&$-1$&$+$\\\hline
\end{tabular}
\caption{A summary of the low energy spectrum of the model considered. The charges include the Standard Model matter content, the $D_4$ family symmetry, the remaining $U(1)_{t_5}$ from the commutant \su{5} descending from $E_8$ orthogonally to the GUT group, and finally the geometric $Z_2$ symmetry.}
\label{les}
\end{table}

Models of the form presented here taken at face-value allow a large number of GUT operators, however we must ensure that all symmetries are respected. This being the case, we find that the tree-level operators found in \tref{oplist}, and constructed from the low energy spectrum summarised in \tref{les}, form the basis for our model, assuming the $D_4$ algebra rules:
\begin{align*}
2\times&2=1_{++}+1_{+-}+1_{-+}+1_{--}\,,\\
1_{a,b}\times&1_{c,d}=1_{ac,bd}\,,\\
\text{with: }\, &a,b,c,d=\pm
\end{align*}
 As well as the expected Yukawas for the quarks and charged leptons, there are also a number of parity violating operators that could lead to dangerous and unacceptable rates of proton decay. However, provided the singlet spectrum is aligned correctly it is possible to avoid unacceptable proton decay rates via dimension 4 operators. It will not be possible to remove all parity violating operators from the spectrum though, and we will be left with $u^cd^cd^c$ operators that may facilitate neutron-antineutron oscillations. It is also possible to remove vector like pairs from the spectrum to insure a low energy matter content similar to the MSSM.
\begin{table}[t!]\centering\small
\begin{tabular}{|c|c|c|c|}\hline
Operator$\to$ type& $D_4$ irrep. &$t_5$ charge&$Z_2$ parity\\\hline
$10_1 10_1 5_b \to QUH$&$1_{++}$&$0$&$1$\\
$10_1 10_2 5_b \to QUH$&$1_{+-}$&$0$&$-1$\\
$10_1 10_3 5_b \to QUH$&$1_{+-}$&$1$&$-1$\\
$10_4 10_1 5_b \to QUH$&$2$&$0$&$1$\\
$10_4 10_2 5_b \to QUH$&$2$&$0$&$-1$\\
$10_4 10_3 5_b \to QUH$&$2$&$1$&$-1$\\\hline
$10_1 \bar{5}_c \bar{5}_d \to QDH$&$1_{++}$&$1$&$1$\\
$10_4 \bar{5}_c \bar{5}_d \to QDH$&$2$&$1$&$1$\\\hline
$10_1 \bar{5}_c \bar{5}_d \to LEH$&$1_{++}$&$1$&$1$\\
$10_4 \bar{5}_c \bar{5}_d \to LEH$&$2$&$1$&$1$\\\hline
$10_1 \bar{5}_c \bar{5}_c \to UDD$&$1_{+-}$&$0$&$-1$\\
$10_2 \bar{5}_c \bar{5}_c \to UDD$&$1_{++}$&$0$&$1$\\
$10_3 \bar{5}_c \bar{5}_c \to UDD$&$1_{++}$&$1$&$1$\\
$10_1 \bar{5}_c \bar{5}_c \to QLD$&$1_{+-}$&$0$&$-1$\\
$10_4 \bar{5}_c \bar{5}_c \to QLD$&$2$&$0$&$-1$\\
$10_1 \bar{5}_c \bar{5}_c \to ELL$&$1_{+-}$&$0$&$-1$\\
$10_4 \bar{5}_c \bar{5}_c \to ELL$&$2$&$0$&$-1$\\\hline
\end{tabular}\caption{List of all trilinear couplings available in the $\text{SU}(5)\times D_4 \times U(1)$ model presented. At tree-level, these operators are not all immediately allowed, since the $D_4$ and $t_5$ symmetries must be respected. \label{oplist}}
\end{table}

\subsubsection{Quark sector}
The up-type quarks have four operators which contribute to the Yukawa matrix. Firstly, we have a tree level top quark coming from the operator $10_110_15_b$, which is the only tree level Yukawa operator found in the Quark and Charged Lepton sectors. The remaining three operators are non-renormalisable operators subject to suppression. We shall assume that the up-type Higgs gets a vacuum expectation value, $\langle H_u\rangle =v_u$. The singlets involved must have vacuum expectation values as summarised in \tref{modeltable2}.
 The following mass terms are generated
\begin{align*}10_110_15_b&\rightarrow y_1v_uQ_3u_3^c\\
10_410_15_b\theta_{a}&\rightarrow y_2v_u(Q_2a_2+Q_1a_1)u_3^c\\
10_410_35_b\theta_{a}\theta_{\beta}&\rightarrow y_3v_u\beta(Q_2a_2+Q_1a_1)u_2^c\\
10_110_35_b\theta_{\beta}&\rightarrow y_4v_u\beta Q_3u^c_2
\end{align*}
giving rise to the  up-quark mass texture
\begin{align*}
M_{u,c,t}&= v_u\left(\begin{array}{ccc}0&y_3a_1\beta&y_2a_1\\0&y_3a_2\beta &y_2a_2\\0&y_4\beta&y_1\end{array}\right).\end{align*}
The lightest generation does not get an explicit mass from this mechanism, but we can expect a small correction to come from non-commutative fluxes or instantons~\cite{Cecotti:2009zf,Aparicio:2011jx,Marchesano:2015dfa}, thus generating a small mass for the first generation.

The down-type quarks contribute a further two operators to the model. These will be symmetric across the righthanded $d^c$ since all three generations are found on the $5_c$ matter curve. We once again assume the Higgs to get a vacuum expectation, $\langle H_d \rangle= v_d$. As before, we also give the singlets a vacuum expectation value: $\langle\theta_{\alpha}\rangle=\alpha$ and $\langle\theta_{\gamma}\rangle=(\gamma_1,\gamma_2)^{T}$. As a result, we get the Yukawa contributions
\begin{align*}
10_1\bar{5}_c\bar{5}_d\theta_{\alpha}&\rightarrow y_{4,i}v_dQ_3d_{i}^c\alpha\\
10_4\bar{5}_c\bar{5}_d\theta_{\gamma}&\rightarrow y_{5,i}v_d(Q_2\gamma_2+Q_1\gamma_1)d_i^c
\end{align*}
and consequently, the down quark mass matrix form
\begin{align*}
M_{d,s,b}&= v_d \left(\begin{array}{ccc} y_{5,1}\gamma_1&y_{5,2}\gamma_1&y_{5,3}\gamma_1\\  y_{5,1}\gamma_2&y_{5,2}\gamma_2&y_{5,3}\gamma_2\\y_{4,1}\alpha&y_{4,2}\alpha&y_{4,3}\alpha\end{array}\right).\end{align*}
However, this mass matrix will be subject to the rank theorem, requiring that there be some suppression factor between the copies of the operator, which we indicate by the second index, $y_{i,j}$.

\subsubsection{Charged Leptons}
The Charged Lepton Yukawas are determined by the same operators as the Down-type quarks, subject to a transpose. As such their mass matrix is as follows:
\begin{align*}
10_1\bar{5}_c\bar{5}_d\theta_{\alpha}&\rightarrow y_{6,i}v_dL_ie_{3}^c\alpha\\
10_4\bar{5}_c\bar{5}_d\theta_{\gamma}&\rightarrow y_{7,i}v_dL_i(e^c_2\gamma_2+e^c_1\gamma_1)\\
M_{e,\mu,\tau}&=v_d \left(\begin{array}{ccc}y_{7,1}\gamma_1&y_{7,1}\gamma_2&y_{6,1}\alpha\\y_{7,2}\gamma_1&y_{7,2}\gamma_2&y_{6,2}\alpha\\y_{7,3}\gamma_1&y_{7,3}\gamma_2&y_{6,3}\alpha\end{array}\right).\end{align*}
The mass relations between charged leptons and down-type quarks will not be constrained to be exact as the operators can be assumed to be localized to different parts of the GUT surface. Once again this is subject to the rank theorem, but will be able to produce a light first generation through other mechanisms.

\subsubsection{Neutrinos}

Over the coming years, all three lepton mixing angles are expected to be measured with increasing precision. A first tentative hint for a value of the CP-violating phase $\delta_{CP} \sim -\pi /2$ has also been reported in global fits \cite{Gonzalez-Garcia:2014bfa,Forero:2014bxa,Capozzi:2013csa}. However the mass squared ordering (normal or inverted), the scale (mass of the lightest neutrino) and nature (Dirac or Majorana) of neutrino mass so far all remain unknown.

On the theory side, there are many possibilities for the origin of light neutrino masses $m_i$ and mixing angles $\theta_{ij}$. Perhaps the simplest and most elegant idea is the classical see-saw mechanism, in which the observed smallness of neutrino masses is due to the heaviness of right-handed Majorana neutrinos \cite{Minkowski:1977sc},
\begin{equation}
m^{\nu}=-m^DM^{-1}_{R}(m^D)^T,
\label{seesaw}
\end{equation}
where $m^{\nu}$ is the light effective left-handed
Majorana neutrino mass matrix (i.e. the physical neutrino mass matrix), $m^D$ is the Dirac mass matrix (in LR convention) and $M_R$ is the (heavy) Majorana mass matrix. Although the see-saw mechanism generally predicts Majorana neutrinos, it does not predict the ``mass hierarchy'', nor does it yield any understanding of lepton mixing. In order to overcome these deficiencies, the see-saw mechanism must be supplemented by other ingredients.
In order to obtain sharp predictions for lepton mixing angles, the relevant Yukawa coupling ratios need to be fixed, for example using vacuum alignment of family symmetry breaking flavons (for reviews see e.g.~\cite{Altarelli:2010gt,Ishimori:2010au,King:2013eh,King:2014nza}).

In F-theory,
neutrinos may admit both Dirac and Majorana mass terms. As such, we would like to use the see-saw mechanism to achieve small neutrino masses via a GUT scale Majorana type mass. Any Dirac type mass comes from an operator of the form
$m_D\sim\theta_{\nu}5_b\bar{5}_c$, while the right-handed Majorana mass terms are of the form $M\theta_{\nu}\theta_{\nu}$. Although we have a non-Abelian $D_4$ family symmetry, the lepton doublets $L$ are in singlet representations (see Table~\ref{les}), so the model offers no opportunity to make predictions for the lepton mixing angles.
 \\

The singlet representations and parities, as detailed in the Appendix A and B, allow us up to nine singlets in this model. Let us then match our right-handed neutrinos to the representations $1_{+-}$ and a doublet, as allowed from our spectrum. This will then give the operators for the Dirac mass:
\begin{align*}
\theta_{\nu_r}5_b\bar{5}_c&\rightarrow y_{8,i}v_u\nu^c_3L_i\\
\theta_{\nu_R}5_b\bar{5}_c\theta_a&\rightarrow y_{9,i}v_u(\nu^c_1a_1+\nu^c_2a_2)L_{i}\\
m_D&= v_u\left(\begin{array}{ccc}y_{9,1}a_1&y_{9,1}a_2&y_{8,1}\\y_{9,2}a_1&y_{9,2}a_2&y_{8,2}\\y_{9,3}a_1&y_{9,3}a_2&y_{8,3}\end{array}\right).
\end{align*}
This Dirac matrix can be shown to be rank two, which will cause our lightest neutrino to be massless. While this is not explicitly ruled out by experiment, a small mass can be generated through some higher order operators from other singlets in the spectrum if required - for example a singlet of the type $1_{--}$ with $+$ parity. This will allow an explicit Dirac type mass, however similar analysis has been done elsewhere ( for example \cite{Karozas:2014aha}), so we omit in depth discussion here.\\

The Majorana terms corresponding to this choice of neutrino spectrum are simply calculated, as one might expect:
\begin{align*}
\theta_{\nu_r}\theta_{\nu_r}&\rightarrow m\nu^c_3\nu^c_3\\
\theta_{\nu_R}\theta_{\nu_R}&\rightarrow M\nu^c_1\nu^c_2\\
\theta_{\nu_r}\theta_{\nu_R}\theta_a&\rightarrow y\nu^c_3\nu^c_2a_2+y\nu^c_3\nu^c_1a_1\\
M_R&= \left(\begin{array}{ccc}0&M&ya_1\\M&0&ya_2\\ya_1&ya_2&m\end{array}\right).
\end{align*}
This may also be allowed corrections via extra singlets, though it will not be needed for this work. \\

The effective neutrino mass can be calculated from the seesaw mechanism via $m_{\nu}=-m_DM_R^{-1}m_D^T$. The resulting mass matrix appears complicated, with elements given in full as:
\begin{align*}
m_{11}=& M  {y_{8,1}}^2+2  {a_1}  {a_2}  {y_{9,1}} ( { m}  {y_{9,1}}-2  {y_{8,1}} y) \\
m_{12}=m_{21}=& M  {y_{8,1}}  {y_{8,2}}-2  {a_1}  {a_2} ( {y_{8,2}} y  {y_{9,1}}- { m}  {y_{9,2}}  {y_{9,1}}+ {y_{8,1}} y  {y_{9,2}})\\
m_{13}=m_{31}=& M  {y_{8,1}}  {y_{8,3}}-2  {a_1}  {a_2} ( {y_{8,3}} y  {y_{9,1}}- { m}  {y_{9,3}}  {y_{9,1}}+ {y_{8,1}} y  {y_{9,3}}) \\
m_{22}=& M  {y_{8,2}}^2+2  {a_1}  {a_2}  {y_{9,2}} ( { m}  {y_{9,2}}-2  {y_{8,2}} y) \\
m_{23}=m_{32}=& M  {y_{8,2}}  {y_{8,3}}-2  {a_1}  {a_2} ( {y_{8,3}} y  {y_{9,2}}- { m}  {y_{9,3}}  {y_{9,2}}+ {y_{8,2}} y  {y_{9,3}}) \\
m_{33}= & M  {y_{8,3}}^2+2  {a_1}  {a_2}  {y_{9,3}} ( { m}  {y_{9,3}}-2  {y_{8,3}} y)
\end{align*}
with an overall scaling of $m_0=v_u^2({M  { m}-2  {a_1}  {a_2} y^2})^{-1}$. \\

In order to extract mixing parameters and mass scales, we will parameterize the matrix in the following way:
\begin{align} X_i=&\frac{y_{8,i}}{y_{8,1}},\; Z_i=\frac{y_{9,i}}{y_{8,1}},\; G=\frac{2a_1a_2}{M}
\end{align}
with $i=1,2,3$, and trivially $X_1=1$. Note that $X_{2,3}$ and $Z_{j}$ are not required to be order one due to the parametrization choice. Let us go a step further, approximating $m\approx M$ and setting $Z_3=0$, then the mass matrix is given by:
\begin{equation}\resizebox{0.9\hsize}{!}{$ m_{\nu}\approx m_0\left(
\begin{array}{ccc}
 G Z_1 (Z_1-2 y)+1 & -G y Z_1 X_2+X_2+G (Z_1-y) Z_2 & X_3-G X_3 y Z_1 \\
 -G y Z_1 X_2+X_2+G (Z_1-y) Z_2 & X_2^2-2 G y Z_2 X_2+G Z_2^2 & X_3 (X_2-G y Z_2) \\
 X_3-G X_3 y Z_1 & X_3 (X_2-G y Z_2) & X_3^2 \\
\end{array}
\right)$}\end{equation}
where:
\be m_0=\frac{v_u^2Mx_1^2}{M^2-Gy^2}\,.\ee
This parametrization allows for comparatively straightforward extraction of mixing parameters. Using Mathematica, we fit the Ratio of mass squared differences in this model to experimental constraints, allowing us to extract a mass scale for the neutrinos while fitting parameters to allow acceptable mixing angles.

 \begin{figure}[!t]
 \begin{flushleft}
\includegraphics[scale=0.285]{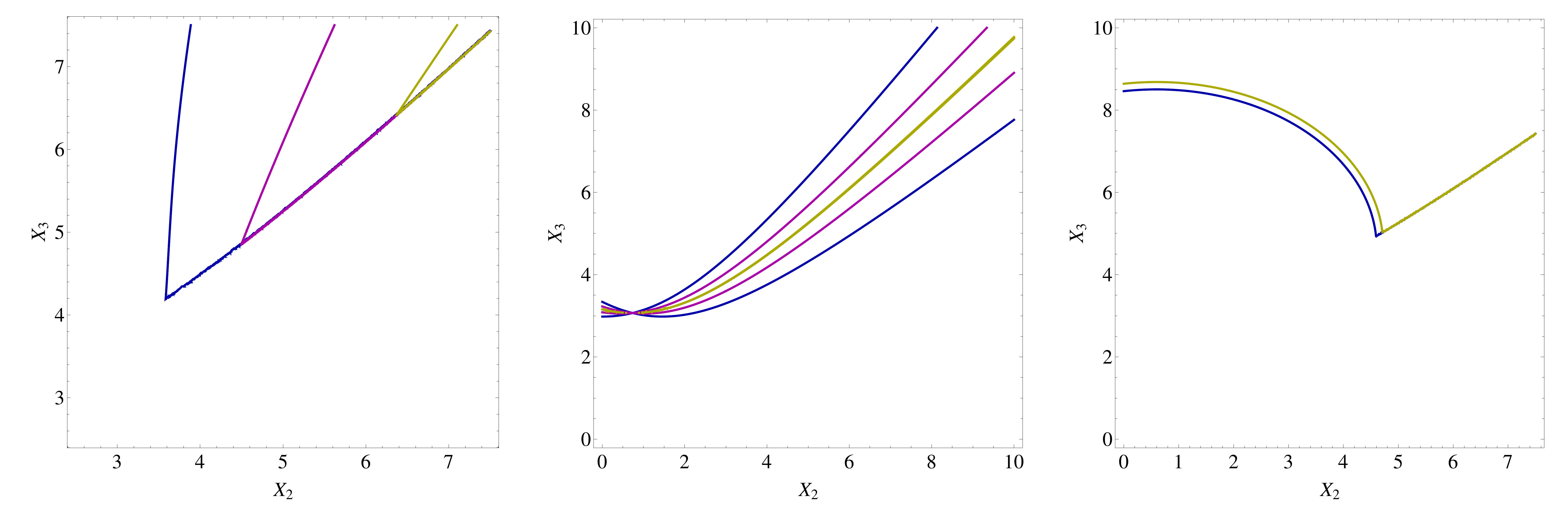}
\end{flushleft}
 \caption{Left: Plot of $\sin^2{\theta_{12}}$ across its $3\sigma$ range (blue-0.270, pink-0.304, yellow-0.344), Center: Plot of $\sin^2{\theta_{23}}$ across its $3\sigma$ range (blue-0.382, pink-0.452, yellow-0.5). Note that the upper bound of $\sin^2{\theta_{23}}$ is $0.643$, but this is not allowed by the model, which permits a maximum of 0.5 for these parameters. Right: Plot of the mass squared difference ratio, R, for its upper and lower bounds of 31.34 (blue) and 34.16 (yellow). For all three plots the parameter space $(X_2,X_3)$ is plotted since these terms should lead the mixing. The remaining parameters are set at values that yield consistent mixing parameters: $(Z_1\approx 2.4, Z_2\approx 4.1, G\approx 0.6, y\approx 0.3)$.\label{nuplts1}}
 \end{figure}

\fref{nuplts1} shows plots of the $3\sigma$ ranges of $\sin^2{\theta_{12}}$, $\sin^2{\theta_{23}}$ and $R=\left|\frac{m_3^2-m_2^2}{m_2^2-m_1^2}\right|$ in the parameter space of $(X_2,X_3)$. This shows that while there are some sharp cutoffs in the parameter space, the key variables can still be allowed. A full simulation of parameters gives, for example:
\begin{align*}
\text{Inputs: }&\left(X_2=4.5,\,X_3=5.7,\,Z_1=2.4,\,Z_2=4.1,\,G=0.6,\,y=0.3\right)\\
\text{Outputs: }&\left(R=33.2,\,\theta_{12}=32.4,\,\theta_{13}=9.07,\,\theta_{23}=39.2\right)
\end{align*}
This also allows us to extract the neutrino masses using the mass differences, which are an implicit input parameter used in calculation of R. We know from the Dirac matrix rank that at this order of operator one neutrino is massless, so then the remaining two masses are (within experimental errors) equal to the square root of the mass differences:
\begin{equation}
m_1=0\ \text{meV}\,,\,m_2=8.66\ \text{meV}\,,\,m_3=50.3\ \text{meV}\,.
\end{equation}
Being at the absolute minimum scale for the neutrino masses, this is automatically compatible with cosmological constraints.

\subsection{$\mu$-Terms}
In this set-up the standard Higgs sector $\mu$-term requires coupling to a singlet in order to cancel the charges under $U(1)_{t_5}$. The most suitable coupling allowed by the singlet sector is a term of the type:
\be \lambda_1 \theta_{\alpha}H_uH_d\,.\ee
As such the $\mu$ term is proportional to the vacuum expectation of the singlet $\theta_{\alpha}$:
\be\mu=\lambda_1\langle\theta_{\alpha}\rangle\ee
Since this singlet couples to the Charged Lepton and the Bottoms quark Yukawa matrices, the resulting vacuum expectation should allow a TeV scale $\mu$-term while not affecting these Yukawas too strongly. Note that since the operators in the Charged Lepton and Bottom quark sectors are non-renormalisable, the coupling should be suppressed by a large mass scale, making this possible. It is also shown in the D-flatness conditions (provided in the appendix) that we have a deal of freedom when choosing the vacuum expectation value for $\theta_{\alpha}$.

A second term of the type:
\be \lambda_2 \theta_{a}\theta_{\gamma}H_uH_d\,\ee
will also contribute to the $\mu$ terms, which is non-renormalisable and should be suppressed by some large mass scale. Refering to the F-flatness conditions and a cursory calculation of this coupling, we see that this contributes proportionally to the product of the vacuum expectations of the $\theta_a$ and $\theta_{\gamma}$ singlets. This again seems acceptable.

\section{Baryon number violation}
\label{baryon}

 \subsection{Proton decay }
It is well known that in the absence of particular types of symmetries
such as R-parity,
  the MSSM  as well as ordinary GUT symmetries are not adequate to ban
rare processes leading
  to  baryon and/or lepton number violation.  Moreover, specific
$SU(5)$  GUT representations include
  additional  states leading to similar drawbacks. Such states are the
  Higgs colour triplets being
  components of the very same fives  containing the up and down SM
Higgs doublets.
  If both  Higgs  fields localise on the same matter curve they generate graphs
  contributing to proton decay from effective operators of the form
$M_{GUT}^{-1}\,QQQL$.
  Since their Yukawa  couplings are expected to be of order one, the
suppression factor $M_{GUT}^{-1}$
  is not sufficient to reduce baryon number violating processes to acceptable rates.

  In F-theory it is possible to  turn on suitable  fluxes so that the
Higgs triplets are removed
   from the low energy spectrum. However even in this case their
associated  Kaluza-Klein modes
  generate the same type of non-renormalisable terms where now the
  suppression factor is replaced by the KK scale $M_{KK}^{-1}$. Since
the $M_{KK}$ mass scale
  is not expected to be substantially larger that the  $M_{GUT}$
scale, one would not
  expect a significant suppression of these operators. It is possible
to achieve further
  suppression however, if the parts of the colour triplet-antitriplet
pair emerge from different
  matter curves so that a direct tree-level mass term is not generated.

In practice, the realistic constructions are more complicated and the
whole issue
of baryon and lepton number violation is more involved. Firstly, as we
have analysed in \sref{gps},
the role of R-parity in this work is played by a $Z_2$ symmetry of
geometric origin which does
not necessarily coincide with the standard R-parity imposed in field
theory supersymmetric models.
Secondly, accompanying symmetries emerging from the $SU(5)_{\perp}$
breaking affix  additional quantum numbers
to the  GUT representations and as such, they imply further
restrictions on  the
superpotential of the effective theory.

We pursue our investigation, elaborating the above for the present model.
Clearly, in order to establish the existence of a proton decay operator,
  we should pay heed to many more factors than in ordinary field theory GUTs,
  such as accompanying symmetries, geometric properties and flux effects.
In the present model, there is a  combination of constraints associated to
 the $D_4$  group, the $Z_2$ discrete symmetry
of geometric origin as well as a $U(1)$ factor that  should be respected.
Although these symmetries eliminate a singificant
number of catastrophic operators, yet there  remain trilinear terms which are
potentially dangerous, which we now discuss.
We start  with the trilinear couplings, which take two forms,
    \ba   10 \cdot\bar 5\cdot\bar 5& \to &Qd^cH_d+QD^cL+e^cLH_d+u^cd^cd^c \\
    10\cdot10\cdot5 &\to& Qu^c H_u+ u^ce^c D^c+QQ D^c
    \ea
   which  in principle, give rise to dimension 5 proton decay  provided  the following coupling exists for the Higgs colour triplet:
    \be   \Phi 5\bar 5 \to  \langle\Phi \rangle D\,D^c \label{DDX}
    \ee
    where $\Phi$ a suitable singlet field acquiring a non-zero vev.
   However, our flux choice eliminates the coloured triplets from Higgs fields (see \tref{modeltable}) and as a result such terms do not exist.


In addition to the above type of operators, there are trilinear
R-parity violating terms that give rise to
proton decay through  similar graphs. Checking \tref{oplist}  one can see
that there is a potentially dangerous baryon violating term,  namely
\be
 10_2 \bar 5_c \bar 5_c \label{dcdcuc}
 \ee
 giving rise to a $u^cd^cd^c$ operator (because of flux effects
 $10_2$ does not contain $Q$, hence the operator $Qd^cL$ does not exist).
Thus, (\ref{dcdcuc})  contributes to proton decay only if analogous
 dimension-four operators from terms of the type $10_i 10_j 5_k$  are
simultaneously present in the superpotential.
In the present model such terms do not exist, hence proton stability is ensured.   
Nevertheless, there are other interesting implications of the above operator that
could be the  low energy  imprint of the present model, which we will now discuss.

\subsection{Neutron-Antineutron oscillations}
\begin{figure}[!t]
  \centering
 {\flushleft \includegraphics[scale=0.6]{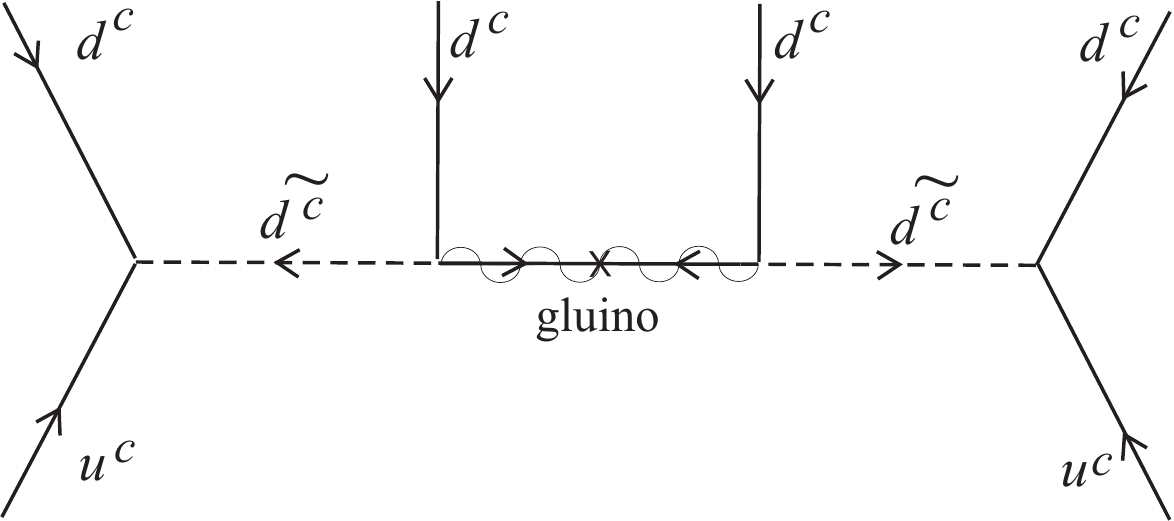}}
{\flushright \includegraphics[scale=0.6]{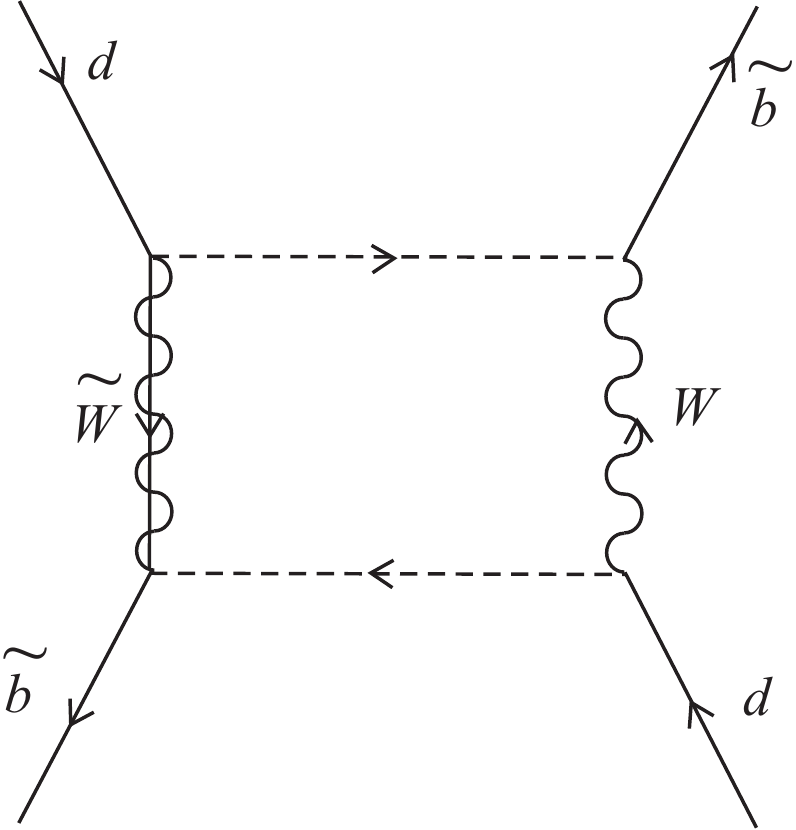}}
  \caption{Feynman graphs for $n-\bar n$ oscillation processes. Top: oscillation via a gluino, Bottom: box-graph process.}
  \label{nnbar}
  \end{figure}

As mention in the previous section, the model presented is free from proton decay at the lowest orders. However, it is subject to operators which are classically considered to be parity violating. Since these operators are all of the type $u^cd^cd^c$, they will instead facilitate neutron-antineutron oscillations. While this is a seldom considered property of GUT models, work has been done to calculate transmission amplitudes of such processes by Mohapatra and Marshak~\cite{Mohapatra:1980qe}
and later on by  Goity and Sher \cite{Goity:1994dq} among others. The contributions to the process are generated  from  tree-level  and box type graphs (see~\cite{Goity:1994dq}, the reviews~\cite{Phillips:2014fgb,Barbier:2004ez}
and references therein), with typical cases shown in \fref{nnbar}.\\

In the paper of Goity and Sher, they argue that one can identify a competitive mechanism, with a fully
calculable transition amplitude, which sets a bound on $\lambda_{dbu}$. This mechanism
is based on the sequence of reactions  $u_{R}d_{R}+d_{L}\rightarrow\tilde{b}_{R}^{*}+d_L
\rightarrow{(\tilde{b}_{L}^{*}+d_L\rightarrow{\bar{d}_L+\tilde{b}_L})}\rightarrow{\bar{d}_L+\bar{u}_{R}\bar{d}_R}$, where the intermediate transition in the parentheses, $\tilde{b}_{L}^{*}+d_L\rightarrow{\bar{d}_L+\tilde{b}_L}$, is due to a W boson and gaugino exchange box diagram . The choice of intermediate bottom squarks is the most favourable one in
order to maximise factors such as $m_{b}^{2}/m_{W}^2$, which arise from the electroweak interactions of
d-quarks in the box diagram (Figure 3). \\

Calculation of the diagram gives the following relation for the decay rate,
\begin{equation}\label{gamma}
\Gamma=-\frac{3g^{4}\lambda^{2}_{dbu}M_{\tilde{b}_{LR}}^{2}m_{\tilde{w}}}{8 \pi^{2}M_{\tilde{b}_{L}}^{4}M_{\tilde{b}_{R}}^{4}}|\psi(0)|^{2}\sum_{j,j'}^{u,c,t}\xi_{jj'}J(M_{\tilde{w}}^{2},M_{W}^{2},M_{u_{j}}^{2},M_{\tilde{u}_{j'}}^{2})
\end{equation}
where the mass term $M_{\tilde{b}_{LR}}$, which mixes $\tilde{b}_{L}$ and $\tilde{b}_{R}$, is given by
$M_{\tilde{b}_{LR}}=Am_{b}$. Here $A$ is the soft SUSY breaking parameter with $A=m_{\tilde{w}}=200GeV$, and $\xi_{jj'}$ is a combination of  CKM matrix parameters,
\be \xi_{jj'}=V_{bu_{j}}V_{u_{j}d}^{\dagger}V_{bu_{j'}}V_{u_{j'}d}^{\dagger}\, \ee
and the $J$ functions are given by:
\be J(m_1,m_2,m_3,m_4)=\sum_{i=1}^{4}\frac{m_{i}^{4}\ln(m_{i}^2)}{\prod_{k\neq{i}}(m_{i}^{2}-m_{k}^{2})}\,.\ee

\begin{figure}[!t]
 \centering
 \includegraphics[scale=1]{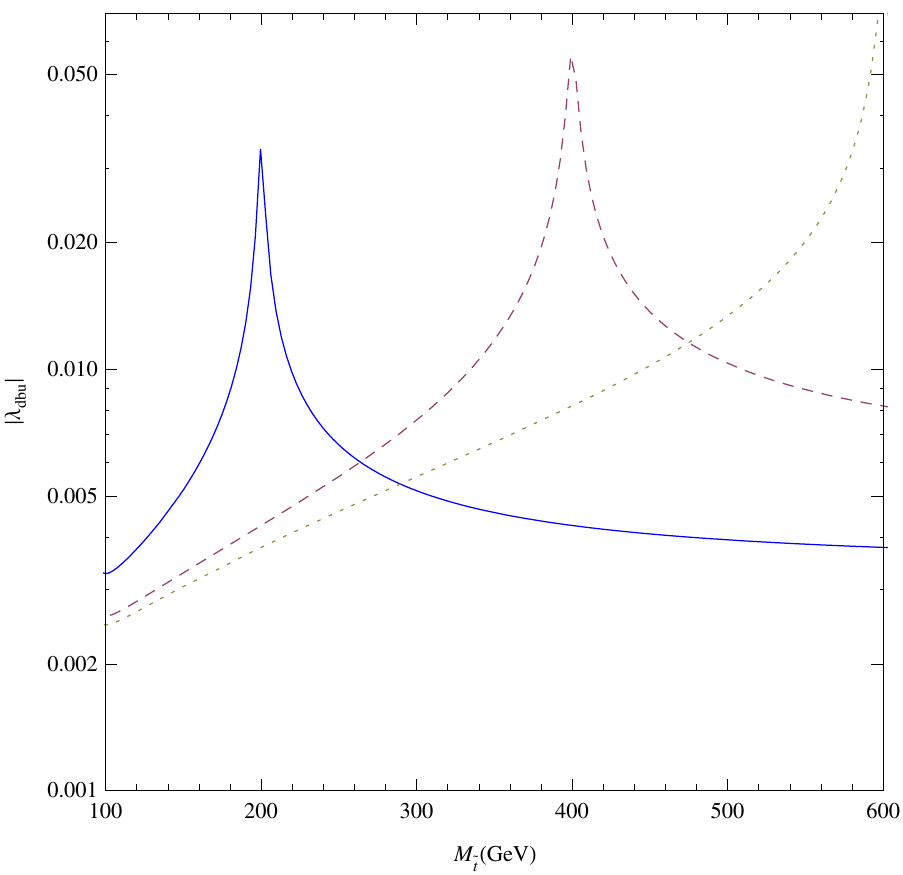}
 \caption{Goity and Sher bounds on $\lambda_{dbu}$. They assumed that up and bottom squark masses are degenerate.
 Blue: $M_{\tilde{u}}=M_{\tilde{c}}=200GeV$, Dashed: $M_{\tilde{u}}=M_{\tilde{c}}=400GeV$, Dotted: $M_{\tilde{u}}=M_{\tilde{c}}=600GeV$. Also we took $M_{\tilde{b}_{L}}=M_{\tilde{b}_{R}}=350GeV$. The peaks corresponds to GIM mechanism effects.}
 \label{GS}
 \end{figure}

\noindent The $n$-$\bar{n}$ oscillation time is $\tau=1/\Gamma$ and the current experimental limits gives, $\tau\gtrsim{10^{8}sec.}$ \cite{Phillips:2014fgb}. Finally $|\psi(0)|$ is
the baryonic wave function matrix element  for three quarks inside
a nucleon. This parameter was calculated to be $|\psi(0)|^{2}=10^{-4}$ and $0.8\times{10^{-4}}GeV^{-6}$ in MIT Bag models\footnote{Goity and Sher used a slightly more stringent bound, $\tau>1.2\times{10^{8}sec.}$ and for the matrix element they took $|\psi(0)|^{2}=3\times{10^{-4}}GeV^{6}$.}. From the experimental limit on the neutron oscillation time we can obtain the bound on $\lambda_{dbu}$. The results depend on CKM parameters and the squarks masses. In Figure \ref{GS} we reproduce the results of Goity and Sher. As we can see the upper bound on $\lambda_{dbu}$ is between 0.005 and 0.1.\\

\begin{figure}[!t]
 \centering
\includegraphics[scale=1]{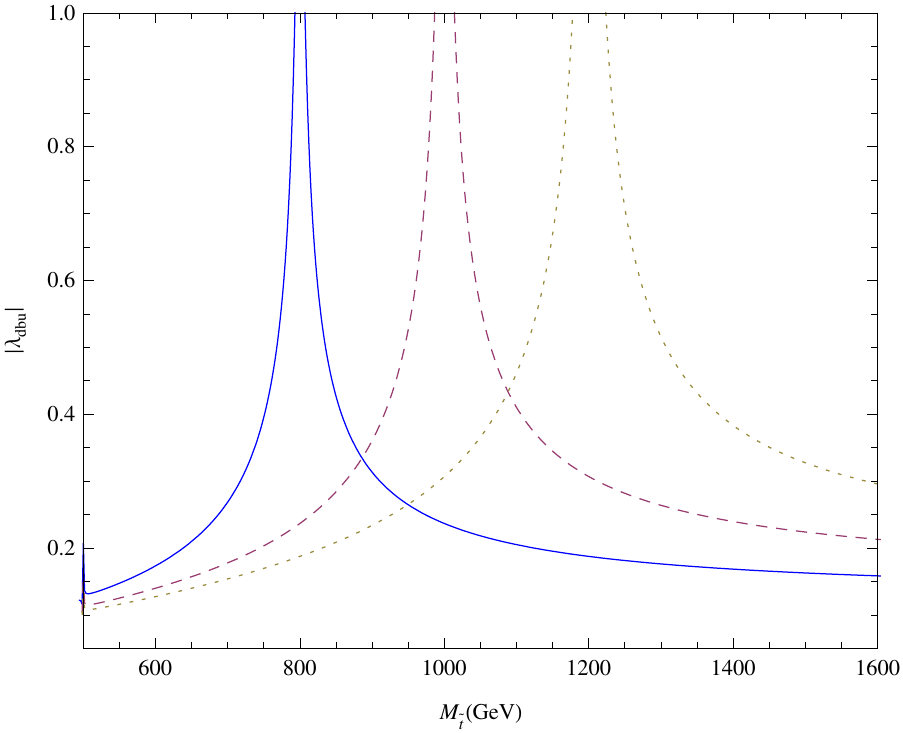}
 \caption{New bounds on $\lambda_{dbu}$ using the latest experimental limits.
 Blue: $M_{\tilde{u}}=M_{\tilde{c}}=800GeV$, Dashed: $M_{\tilde{u}}=M_{\tilde{c}}=1000GeV$, Dotted: $M_{\tilde{u}}=M_{\tilde{c}}=1200GeV$. Also we use the following values for the other parameters: $M_{\tilde{b}_{L}}=M_{\tilde{b}_{R}}=500GeV$, $\tau=10^{8}sec.$ and $|\psi(0)|=0.9\times{10^{-4}}GeV^{-6}$.}
 \label{sn1}
 \end{figure}

Next we use the Equation \eqref{gamma} to recalculate the bounds on $\lambda_{dbu}$ with the latest experimental results for the SUSY mass parameters. In Figure \ref{sn1} the curves correspond to squark masses of 800, 1000 and 1200GeV (Blue, dashed and dotted accordingly). As we can see the value of $\lambda_{dbu}$ lies between 0.1 and $\sim$ 0.5 for stop mass between 500 and 1600GeV, neglecting GIM effects.\\

In F-theory there is an associated  
wavefunction~\cite{Heckman:2008qa}-\cite{Font:2012wq}
to the state residing on each  matter curve and it can  be determined  
by solving the corresponding
equations of motion~\cite{Beasley:2008dc}.  The solutions show that  
each wavefunction is peaked
along the corresponding matter curve.   Yukawa couplings are formed at  
the point of intersection
of three matter curves where the corresponding wavefunctions  overlap.  
To estimate the corresponding
Yukawa coupling  we need to perform  an integration over  the three  
overlapping wavefunctions of the
corresponding states participating in the trilinear coupling. Taking  
into account mixing effects this particular
coupling is  estimated to be of the order $\lambda_{dbu}\le 10^{-1}$.  
From the figure it can be observed that
recent $n-\bar n$ oscillation bounds on  $\lambda_{dbu}$  are  
compatible  with such values.


\section{Conclusions}
\label{conclusions}
In this work an F-theory derived \su{5} model was constructed, with the implications of the arising non-Abelian familiy symmetry being considered, following from work in \cite{Antoniadis:2013joa} and \cite{Karozas:2014aha}. Using the spectral cover formalism, assuming a point of $E_8$ enhancement descending to an \su{5} GUT group, the corresponding maximal symmetry (also \su{5}) should reduce down to  a subgroup of the Weyl group, $S_5$. In this paper we  derive the conditions on the spectral cover equation in the case of the non-abelian discrete group  $D_4$, which was assumed to play the role of a family symmetry.  A novel geometric symmetry was also employed to produce an R-parity-like $Z_2$ symmetry. The combined effect of this framework on the effective field theory has been examined, and the resulting model shown to exhibit parity violation in the form of neutron-antineutron oscillations, while being free from dangerous proton decay operators. The experimental constraints on this interesting process have been calculated, using current data on the masses of supersymmetric partners. Detection of such baryon-violating processes, without proton decay, serve as a potential smoking gun for this type of model.

The physics of the neutrino was also considered, and it was shown that at lowest orders this model predicts a massless first generation neutrino. Correspondingly, the masses of the two other generations then equate to the mass differences from experiment, with the hierarchy being normal ordered. The mixing angles were also probed numerically, with results that are consistent with large mixing in the neutrino sector and a non-zero reactor mixing angle.

In conclusion F-theory model building  predicts in a natural way the coexistence of GUT models with non-Abelian discrete 
symmetry extensions. The reach symmetry content following from the decomposition of the $E_8$ covering group and the  
geometric symmetries emerging from the internal manifold structure are sufficient to incorporate successful non-Abelian 
groups  which  have already been proposed  in phenomenological constructions during the last decade. The distinct role
 of the discrete groups as family symmetries occurs naturally in the F-theory constructions.
Moreover, the theory provides powerful tools  to get an effective field theory with definite predictions.

\section{Acknowledgments}
\label{acknowledgments}
SFK acknowledges partial support from the  European Union FP7  ITN INVISIBLES
 (Marie Curie Actions, PITN- GA-2011- 289442). AM is supported by an STFC studentship.
 GKL  acknowledges partial support  from the European Union (European Social Fund - ESF) and
Greek national funds through the Operational Program ``Education and
Lifelong Learning'' of the National Strategic Reference Framework
(NSRF) - Research Funding Program: ``ARISTEIA''. Investing in the
society of knowledge through the European Social Fund.


\newpage
\appendix
\section{Irreducible representations of $D_4$}
\begin{figure}[b!]\centering
\includegraphics[scale=0.6]{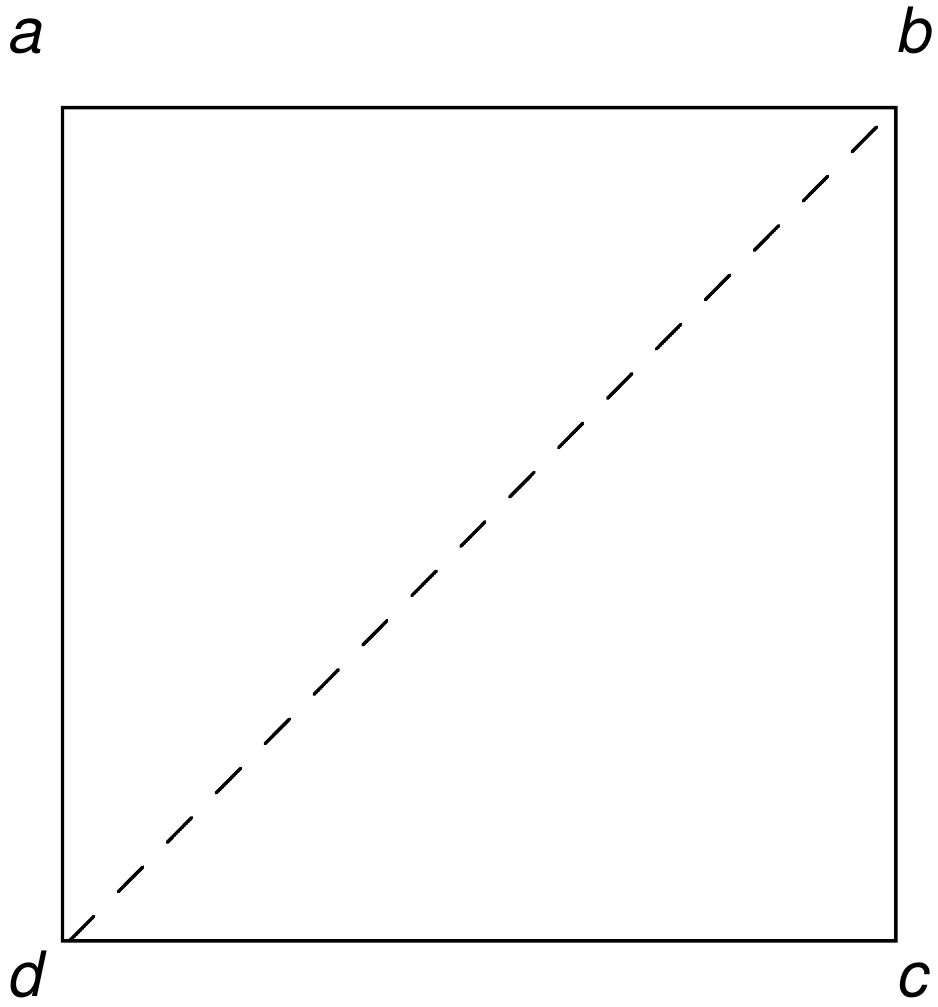}
\caption{A physical representation of the symmetry group $D_4$. The dashed line shows a possible reflection symmetry, while it also has a rotational symmetry if rotated by $\frac{n\pi}{2}$.\label{square}}
\end{figure}

Since we have four weights related, the representation of the 10s of the GUT group will be quadruplets of $D_4$: $(t_1,t_2,t_3,t_4)^{\text{T}}$. Physically we may take each of these weights to represent a corner of a square (or an equivalent interpretation). These weights will transform in this representation such that the two generators required to describe all possible transformations are equivalent to a rotation about the center of the square of $\frac{\pi}{2}$ and a reflection about a line passing through the center - say the diagonal running between the top right and bottom left corners (see Figure \ref{square}).
\\

The two generators are:
\begin{align}
a=&\left(\begin{array}{cccc}0&0&0&1\\1&0&0&0\\0&1&0&0\\0&0&1&0\end{array}\right)\,,\\
b=&\left(\begin{array}{cccc}1&0&0&0\\0&0&0&1\\0&0&1&0\\0&1&0&0
\end{array}\right)\,.\end{align}
These generators must obey the general conditions for dihedral groups, which for $D_4$ are:
\begin{align}
a^4=b^2=&\text{I}\\
b\cdot a\cdot b=&a^{-1}
\end{align}
It is trivial to see that these conditions are obeyed by our generators. In order to obtain the irreducible representations we should put this basis into block-diagonal form, which is achieved by applying the appropriate unitary matrices.\\

Since $D_4$ is known to have a two-dimensional irreducible representation, we might assume that our four-dimensional case can be taken to a block diagonal form including either a doublet and two singlets or two doublets via a unitary transformation.\\

If we initially assume two doublets, then we may put some conditions on our unitary matrix:
\begin{align}
A'=&U\cdot A\cdot U^{\text{T}}=\left(\begin{array}
{cccc}-&-&0&0\\-&-&0&0\\0&0&-&-\\0&0&-&-\end{array}\right)\\
B'=&U\cdot B\cdot U^{\text{T}}=\left(\begin{array}
{cccc}-&-&0&0\\-&-&0&0\\0&0&-&-\\0&0&-&-\end{array}\right)\\
\text{I}=& U\cdot U^{\text{T}}\,.
\end{align}
If we make use of these conditions, there are a number of equivalent solutions for $U$, one of which is:
\begin{equation}U=\frac{1}{\sqrt{2}}
\left(\begin{array}
{cccc}1&0&1&0\\0&1&0&1\\1&0&-1&0\\0&1&0&-1
\end{array}\right)\,.
\end{equation}
This matrix will give a block diagonal form for the generators. Explicitly this is:
\begin{align}
A'=&\left(\begin{array}{cccc}0&1&0&0\\1&0&0&0\\0&0&0&-1\\0&0&1&0\end{array}\right)\,,\\
B'=&\left(\begin{array}{cccc}1&0&0&0\\0&1&0&0\\0&0&1&0\\0&0&0&-1\end{array}\right)\,, \\
\left(\begin{array}{c}t_1\\t_2\\t_3\\t_4\end{array}\right)&\rightarrow\frac{1}{\sqrt{2}}\left(\begin{array}{c}t_1+t_3\\t_2+t_4\\t_1-t_3\\t_2-t_4\end{array}\right)\,.
\end{align}
A cursory examination reveals that the conditions for $D_4$ are still fulfilled by this new basis, and it would seem that at a minimum we have two doublets of the group. However we shall now examine if one of the doublets decomposes to two singlets.

The upper block of the $B'$ generator takes the form of the identity, so we might suppose that the first of our two doublets could decompose into two singlets. Using the same conditions as for the four-dimensional starting point, which can be enforced on the two-dimensional case, we can find easily that:
\begin{align}
V=&\frac{1}{\sqrt{2}}\left(\begin{array}{cc}1&1\\1&-1\end{array}\right)\\
A''=&\left(\begin{array}
{cc}1&0\\0&-1\end{array}\right)\\
B''=&\left(\begin{array}{cc}1&0\\0&1\end{array}\right)\\
\frac{1}{\sqrt{2}}\left(\begin{array}{c}t_1+t_3\\t_2+t_4\end{array}\right)&\rightarrow\frac{1}{2}\left(\begin{array}{c}t_1+t_2+t_3+t_4\\t_1-t_2+t_3-t_4\end{array}\right)
\end{align}
It would seem then in this case that the four-dimensional representation of $D_4$ can be reduced to a doublet and two singlets forming an irreducible representation. The type of the singlets can be determined by examination of the conjugacy classes of the group, which reveals that the upper singlet is of the type $1_{++}$, while the lower is $1_{+-}$. Table 2 summarising the representations of the tens.

\subsection{$D_4$ representations for GUT group Fundamental representation}
The roots of the five-curves can also be described in terms of the roots:
\be t_i+t_j=0\,\forall i\neq j\,.\ee
which gives a total of ten solutions, though these will be  related by the discrete group. Under the $D_4$ symmetry, we can see trivially that since the weight $t_5$ is chosen to be the invariant root, all the roots corresponding to the fives of the form $t_i+t_5$ will transform separately to the $i,j\neq5$ roots. In fact, these will form a doublet and two singlets:  $1_{++}$ and  $1_{+-}$.\\

The remaining six roots of $P_5$ can be constructed into a sextet:
\be R_6=\left(\begin{array}{c}t_1+t_3\\t_2+t_4\\t_1+t_2\\t_3+t_4\\t_1+t_4\\t_2+t_3\end{array}\right)\,.\ee
By construction, we have arranged that the array manifestly has block diagonal generators, $A$ and $B$, such that the first two lines have generators:
\be A=\left(\begin{array}{cc}0&1\\1&0\end{array}\right)\,\,\,B=\left(\begin{array}{cc}1&0\\0&1\end{array}\right)\,.\ee
We can again refer to the previous results to see that this reduces to two singlets: $1_{++}$ and  $1_{+-}$.\\

The remaining quadruplet has generators:
\be A=\left(\begin{array}{cccc}0&0&1&0\\0&0&0&1\\0&1&0&0\\1&0&0&0\end{array}\right)\,\,\,B=\left(\begin{array}{cccc}0&0&0&1\\0&0&1&0\\0&1&0&0\\1&0&0&0\end{array}\right)\,,
\ee
which we can block diagonalise using the unitary matrix:
\be U=\frac{1}{\sqrt{2}}\left(
\begin{array}{cccc}
 1 & 1 & 0 & 0 \\
 0 & 0 & 1 & 1 \\
 -1 & 1 & 0 & 0 \\
 0 & 0 & 1 & -1 \\
\end{array}
\right)\,.\ee
This gives two blocks, which are distiniguished principally by their $A$ generators:
\be A'=\left(
\begin{array}{cccc}
 0 & 1 & 0 & 0 \\
 1 & 0 & 0 & 0 \\
 0 & 0 & 0 & -1 \\
 0 & 0 & 1 & 0 \\
\end{array}
\right)\,\,\,B'=\left(
\begin{array}{cccc}
 0 & 1 & 0 & 0 \\
 1 & 0 & 0 & 0 \\
 0 & 0 & 0 & 1 \\
 0 & 0 & 1 & 0 \\
\end{array}
\right)\,.\ee
The upper block can be further diagonalised to yield two singlets, using the unitary matrix:
\begin{align} V_u=\frac{1}{\sqrt{2}} \left(
\begin{array}{cc}
 1 & 1 \\
 1 & -1 \\
\end{array}
\right)\,,\\
A''_u=B''_u=\left(
\begin{array}{cc}
 1 & 0 \\
 0 & 1 \\
\end{array}
\right)\,,
\end{align}
which, after consulting a character table for the group, returns two singlets of the type $1_{++}$. \\

The lower block can be rotated into the usual doublet basis by the matrix:
\be V_d=\frac{1}{\sqrt{2}} \left(
\begin{array}{cc}
 1 & 1 \\
 -1 & 1 \\
\end{array}
\right)\,.\ee
The full set of states arising from the five-curves is given in Table \ref{Irreps5}.

\subsection{$D_4$ Representations for GUT Group Singlet Spectrum}
The singlets in F-theory correspond to differences of weights of the perpendicular group:
\begin{align*}
\pm&(t_i-t_j)\,\,\,\forall i\neq j\,.
\end{align*}
As such in the case of an $SU(5)$ GUT group we have a total of 20 possible singlets allowed on the GUT surface. Note that four of the singlets have no weight. In the case where four of the roots are related by a $D_4$, the singlets can be considered to split into two different sets:
\begin{align*}
\pm&(t_i-t_j)=0\,,\\
\pm&(t_i-t_5)=0\,,\\
&i\neq j\,.
\end{align*}
This is obvious given that we consider $t_5$ not to transform with the $D_4$ action.
\subsubsection{$\pm(t_i-t_5)$}
In the event $\bm{t_i-t_5}$ is considered we can essentially ignore the $t_5$ weight, since it doesn't transform. Then we can immediately refer to the known result for decomposing the $10$s of the GUT group:
\begin{align*}\left(
\begin{array}{c}
t_1-t_5\\t_2-t_5\\t_3-t_5\\t_4-t_5
\end{array}\right)\to 1_{++}^{-t_5}+1_{+-}^{-t_5}+2^{-t_5}
\end{align*}
The diagonalising matrix is:
\begin{align*}
U=\frac{1}{2}\left(
\begin{array}{cccc}
 1 & 1 & 1 & 1 \\
 1 & -1 & 1 & -1 \\
 \sqrt{2} & 0 & -\sqrt{2} & 0 \\
 0 & \sqrt{2} & 0 & -\sqrt{2} \\
\end{array}
\right)
\end{align*}\\

For $\bm{t_5-t_i}$, we expect a similar decomposition by symmetry. However, if we decompose to the same generators as the $\bm{t_i-t_5}$ case, then the $t_i$ charges are negative.

\subsubsection{$\pm(t_i-t_j)$}
The $t_5$-free singlet combinations fill out 12 combinations. In the \textquotedblleft traditional " interpretation of
a  monodromy group in F-theory, these would all be weightless. I.e. because we identify $t_i$ (with $i=1,2,3,4$) under our monodromy group action they would all have $t_i-t_i=0$. \\

However, in the case that we have a non-Abelian group such as $D_4$ the weights are not directly identified. In this case the irreducible representations appear to be important. We can treat these in a few \textquotedblleft clusters", which will simplify block diagonalising. Firstly:

\begin{align*}
\left(\begin{array}{c} t_1-t_3\\t_2-t_4\\t_3-t_1\\t_4-t_2
\end{array}\right)&\to\left(
\begin{array}{c}
 t_4-t_2 \\
 t_1-t_3 \\
 t_1+t_3 \\
 t_2+t_4
\end{array}
\right)
\end{align*}
The upper-block is manifestly a doublet of the type already encountered in other part of the spectrum, while the lower part can be rotate into a basis with a trivial singlet and a non trivial singlet: $1_{++}$ and $1_{+-}$. \\

\begin{table}[t!]\centering
\begin{tabular}{|c|c|c|c|}\hline
$D_4$ rep. & $t_5$& $t_i$&Type\\\hline
$1_{++}$&$-1$  & $t_1+t_2+t_3+t_4$ &$\theta_{\alpha}$\\
$1_{+-}$&  $-1$& $t_1-t_2+t_3-t_4$ &$\theta_{\beta}$ \\
$2$&  $-1$& $\left(\begin{array}{c} t_4-t_2\\t_1-t_3\end{array}\right)$ &$\theta_{\gamma}$ \\
$1_{++}$& $+1$ &  $-t_1-t_2-t_3-t_4$ &$\theta_{\alpha}'$\\
$1_{+-}$& $+1$ & $-t_1+t_2-t_3+t_4$ &$\theta_{\beta}'$  \\
$2$&  $+1$& $\left(\begin{array}{c} t_2-t_4\\t_3-t_1\end{array}\right)$ &$\theta_{\gamma}'$\\
$1_{++}$& $0$ & $t_1+t_2+t_3+t_4$&$\theta_1$ \\
$1_{+-}$&$0$  & $t_1-t_2+t_3-t_4$ &$\theta_2$\\

$1_{+-}$&$0$  & $-t_1+t_2-t_3+t_4$ &$\theta_2$\\
$1_{--}$& $0$ & $-t_1-t_2-t_3-t_4$ &$\theta_3$ \\
$2$&$0$  &  $\left(\begin{array}{c} t_2-t_4\\t_3-t_1\end{array}\right)$&$\theta_4$\\
$1_{+-}$&$0$  &  $t_1-t_2+t_3-t_4$ &$\theta_2$\\
$1_{--}$&$0$  &  $t_1+t_2+t_3+t_4$ &$\theta_3$\\
$2$& $0$ & $\left(\begin{array}{c} t_4-t_2\\t_1-t_3\end{array}\right)$ &$\theta_4$\\\hline
\end{tabular}\caption{The complete list of the irreducible representations of $D_4$ obtained by block diagonalizing the singlets of the GUT group. Each of these GUT singlets is duly labeled $\theta_i$ to classify them, since some appear to be in some sense degenerate.\label{Irreps1}}
\end{table}

The remaining weight combinations have a symmetry under exchange of $i\to-i$ that allows them to be decomposed into two sets:
\be \pm\left(\begin{array}{c}
t_1-t_2\\t_2-t_3\\t_3-t_4\\t_4-t_1
\end{array}\right)\ee
These can be decomposed into a doublet and two singlets by:
\be U=\frac{1}{2}\left(
\begin{array}{cccc}
 0 & \sqrt{2} & 0 & -\sqrt{2} \\
 -\sqrt{2} & 0 & \sqrt{2} & 0 \\
 1 & -1 & 1 & -1 \\
 1 & 1 & 1 & 1 \\
\end{array}
\right)\ee
The interesting result here is that the singlets are of the types: $1_{+-}$ and $1_{--}$, which is unique to our singlet sector. A complete list of the singlet spectrum is given in \tref{Irreps1}\\

\subsection{Basic Galois Theory}

According to Galois theory if $\mathcal{L}$ is the splitting field of a separable polynomial $P\in{\mathcal{F}[x]}$, then the Galois group $Gal(\mathcal{L}/\mathcal{F})$ is associated with the permutations of the roots of $P$. Let $P$ has degree $n$. Then in $\mathcal{L}[x]$ we can write the $P$ as the product

\begin{equation}
P(x)=c(x-t_{1})\ldots(x-t_{n})
\end{equation}

\noindent where $c\neq{0}$ and the roots $t_{1},\ldots{t_{n}}\in{\mathcal{L}}$ are distinct. In this situation we get a map

\begin{equation*}
Gal(\mathcal{L}/\mathcal{F})\rightarrow{S_{n}}
\end{equation*}

\noindent which is a one-to-one group homomorphism.  Important r$\hat{o}$le in the determination of the Galois group of a polynomial plays the discriminant, which is a symmetric function of the roots $t_{i}$. The discriminant $\Delta(P)\in{\mathcal{F}}$ of a (monic) polynomial $P\in{\mathcal{F}[x]}$ with $P=(x-t_{1})\ldots{(x-t_{n})}$ in a splitting field $\mathcal{L}$ of $P$ is

\begin{equation}
\Delta(P)=\prod_{i<j}(t_{i}-t_{j})^{2}.
\end{equation}

\noindent Another useful object is the square root of the discriminant:

\begin{equation}
\sqrt{\Delta(P)}=\prod_{i<j}(t_{i}-t_{j})\quad{\in{\mathcal{L}}}.
\end{equation}

\noindent Note that while $\Delta$ is uniquely determined by $P$, the above square root depends on how the roots are labeled. It is obvious that the $\sqrt{\Delta(P)}$ controls the relation between $Gal(\mathcal{L}/\mathcal{F})$ and the alternating group $A_{n}\subset{S_{n}}$. More precisely, the image of $Gal(\mathcal{L}/\mathcal{F})$ lies in $A_{n}$ if and only if $\sqrt{\Delta(P)}\in{\mathcal{F}}$ (i.e., $\Delta(P)$ is the square of an element of $\mathcal{F}$). In our case we deal with a fourth degree polynomial corresponding to
the spectral surface $C_4$, hence our starting point is $S_4$ and $A_4$.

To reduce further the $S_{4}/A_{4}$ down to their subgroups ($D_{4}$, $Z_{4}$ and $V_{4}$) we need the service of the so called \emph{resolvent cubic} of $P$

\begin{equation}
R_{3}=(x-x_{1})(x-x_{2})(x-x_{3})
\end{equation}

\noindent where now the $x_{i}$'s are symmetric polynomials of the roots with

\begin{equation}
x_{1}=t_{1}t_{2}+t_{3}t_{4},\quad{x_{2}=t_{1}t_{3}+t_{2}t_{4}},\quad{x_{3}=t_{3}t_{2}+t_{1}t_{4}}.
\label{roots1}
\end{equation}

\noindent A permutation of the indices carries $x_{1}$ to one of the three polynomials $x_{i}$, i=1,2,3. Since $S_{4}$ has order 24, the stabilizer of $x_{1}$ is of order 8, it is one of the three dihedral groups $D_{4}$. Also, $\Delta{(R_{3})}=\Delta{(P)}$, so when $P$ is separable so is $R_{3}$.  Using the discriminant and the reducibility of the cubic resolvent we can correlate the groups $S_{4}, D_{4},Z_{4}, A_{4}$ and $V_{4}$ with the Galois group of a quartic irreducible polynomial. The analysis above with respect to $\Delta(P)$ and $R_{3}$ is summarized in \tref{Galoisgroup}.

\begin{table}[t]\centering\begin{tabular}{|c|c|c|}\hline
 $\Delta(P)$ & $R_{3}$ in $\mathcal{F}$ & $Gal(\mathcal{L}/\mathcal{F})$ \\\hline
  $\neq{\square}$ & irreducible & $S_{4}$ \\
  $=\square$ & irreducible & $A_{4}$ \\
  $\neq{\square}$ & reducible & $D_{4}$ or $Z_{4}$\\
  $=\square$ & reducible & $V_{4}$\\
  \hline
\end{tabular}
\caption{The Galois groups for the various cases of the discriminant and the reducibility of the cubic resolvent $R_{3}$.}\label{Galoisgroup}
\end{table}

\pagebreak

\section{Flatness Conditions}

In order to obtain a realistic model we use the SU(5) singlets which acquire VEV's . Any such VEV's should be consistent with F and D flatness conditions. Singlets spectrum in F-Theory is described by the equation

\begin{displaymath}
\prod_{i\neq{j}}(t_{i}-t_{j})=0
\end{displaymath}

 \noindent  where the product is the discriminant of the spectral cover polynomial. By calculating the discriminant  using the $b_{1}=0$ constraint as  well as the splitting options we end up with the following equation

 \begin{align}
\nonumber &a_0 a_2^3 a_7^2 \left(-a_7^3 \kappa -a_2 a_7^2 \lambda  \mu ^2+2 a_0 a_2^3 \mu ^4+a_2 a_7^2 \mu \right){}^2\\\nonumber
 &\left(256 a_0^2 a_7^3 a_2^2 \kappa ^3+128 a_0
   a_7^4 a_2 \kappa ^2 \lambda ^2+144 a_0^2 a_7^2 a_2^3 \kappa ^2 \lambda  \mu ^2+27 a_0^3 a_2^5 \kappa ^2 \mu ^4+192 a_0^2 a_7^2 a_2^3 \kappa ^2 \mu +16
   a_7^5 \kappa  \lambda ^4\right.\\\nonumber
   &\left. +4 a_0 a_7^3 a_2^2 \kappa  \lambda ^3 \mu ^2 -18 a_0^2 a_7 a_2^4 \kappa  \lambda  \mu
   ^3-144 a_0 a_7^3 a_2^2 \kappa  \lambda
    -6 a_0^2 a_7 a_2^4 \kappa  \mu ^2-4 a_7^4 a_2 \lambda ^3-a_0 a_7^2 a_2^3 \lambda ^2 \mu ^2\right.\\
    &\left.+18 a_0 a_7^2 a_2^3
   \lambda  \mu -80 a_0 a_7^3 a_2^2 \kappa  \lambda ^2 \mu +4 a_0^2 a_2^5 \mu ^3+27 a_0 a_7^2 a_2^3\right)=0
 \end{align}

 \noindent As we can see we have nine factors, four of which correspond to a negative parity (the $a_{0}$ factor, the double factor $ \left(-a_7^3 \kappa -a_2 a_7^2 \lambda  \mu ^2+2 a_0 a_2^3 \mu ^4+a_2 a_7^2 \mu \right)$ and $256 a_0^2 a_7^3 a_2^2 \kappa ^3+\dots$).

\subsection{\textit{F}-flatness}

In general the Superpotential for the massless singlet fields ($\theta_{ij}\equiv\theta_{t_{i}-t_{j}}$) is given by

\begin{equation}
\mathcal{W}=\mu_{ijk}\theta_{ij}\theta_{jk}\theta_{ki}
\end{equation}
\noindent and the F-flatness conditions are given by :
\begin{equation}
\frac{\partial\mathcal{W}}{\partial\theta_{ij}}=\mu_{ijk}\theta_{jk}\theta_{ki}=0.
\end{equation}

\noindent For the model presented in the main text, the singlet superpotential reads

\begin{align}
\nonumber\mathcal{W}_{\theta}&=\mu_{1}\theta_{1}\theta_{\alpha}\theta_{\alpha}'+\mu_{2}\theta_{1}\theta_{\beta}\theta_{\beta}'+\mu_{3}\theta_{1}\theta_{\gamma}\theta_{\gamma}'
+\mu_{4}\theta_{3}\theta_{\gamma}\theta_{\gamma}'
\\\nonumber
&+\lambda_{1}\theta_{4}\theta_{\gamma}\theta_{\alpha}'
+\lambda_{2}\theta_{4}\theta_{\gamma}'\theta_{\alpha}+\lambda_{3}\theta_{4}'\theta_{\gamma}'\theta_{\beta}+\lambda_{4}\theta_{4}'\theta_{\gamma}\theta_{\beta}'\\
&+\lambda_{5}\theta_{2}\theta_{\alpha}\theta_{\beta}'+\lambda_{6}\theta_{2}\theta_{\alpha}'\theta_{\beta}+\lambda_{7}\theta_{2}\theta_{4}\theta_{4}'
\end{align}

 \noindent where  all the singlets have positive parity except the $\theta_{\beta}$, $\theta_{\beta}'$, $\theta_{2}$ and $\theta_{4}'$. Here with $\theta_{4}$ we mean the $\theta_{a}$ ($\theta_{4}'$\emph{corresponds to} $\nu_{R}$).

 Minimization of the superpotential leads to the following equations:
\begin{align*}
\frac{\partial\mathcal{W}}{\partial\theta_{1}}&=\mu_{1}\theta_{\alpha}\theta_{\alpha}'+\mu_{2}\theta_{\beta}\theta_{\beta}'+\mu_{3}\theta_{\gamma}\theta_{\gamma}'=0\\
\frac{\partial\mathcal{W}}{\partial\theta_{2}}&=\lambda_{5}\theta_{\alpha}\theta_{\beta}'+\lambda_{5}\theta_{\alpha}'\theta_{\beta}+\lambda_{7}\theta_{4}\theta_{4}'=0\\
\frac{\partial\mathcal{W}}{\partial\theta_{3}}&=\mu_{4}\theta_{\gamma}\theta_{\gamma}'=0\\
\frac{\partial\mathcal{W}}{\partial\theta_{4}}&=\lambda_{1}\theta_{\gamma}\theta_{\alpha}'+\lambda_{2}\theta_{\gamma}'\theta_{\alpha}'+\lambda_{7}\theta_{2}\theta_{4}'=0\\
\frac{\partial\mathcal{W}}{\partial\theta_{4}'}&=\lambda_{3}\theta_{\gamma}'\theta_{\beta}+\lambda_{4}\theta_{\gamma}\theta_{\beta}'+\lambda_{7}\theta_{2}\theta_{4}=0\\
\frac{\partial\mathcal{W}}{\partial\theta_{\alpha}}&=\mu_{1}\theta_{1}\theta_{\alpha}'+\lambda_{2}\theta_{4}\theta_{\gamma}'+\lambda_{5}\theta_{2}\theta_{\beta}'=0\\
\frac{\partial\mathcal{W}}{\partial\theta_{a}'}&=\mu_{1}\theta_{1}\theta_{\alpha}+\lambda_{1}\theta_{4}\theta_{\gamma}+\lambda_{6}\theta_{\beta}\theta_{2}=0\\
\frac{\partial\mathcal{W}}{\partial\theta_{\beta}}&=\mu_{2}\theta_{1}\theta_{\beta}'+\lambda_{3}\theta_{4}'\theta_{\gamma}'+\lambda_{6}\theta_{2}\theta_{\alpha}'=0\\
\frac{\partial\mathcal{W}}{\partial\theta_{\beta}'}&=\mu_{2}\theta_{1}\theta_{\beta}+\lambda_{4}\theta_{4}'\theta_{\gamma}+\lambda_{5}\theta_{2}\theta_{\alpha}=0\\
\frac{\partial\mathcal{W}}{\partial\theta_{\gamma}}&=\mu_{3}\theta_{1}\theta_{\gamma}'+\mu_{4}\theta_{3}\theta_{\gamma}'+\lambda_{1}\theta_{4}\theta_{\alpha}'+\lambda_{4}\theta_{4}'\theta_{\beta}'=0\\
\frac{\partial\mathcal{W}}{\partial\theta_{\gamma}'}&=\mu_{3}\theta_{1}\theta_{\gamma}+\mu_{4}\theta_{3}\theta_{\gamma}+\lambda_{2}\theta_{4}\theta_{\alpha}+\lambda_{3}\theta_{4}'\theta_{b}=0
\end{align*}

 \noindent As we can see we have a system of  11-equations. Solving the system with the requirements $\langle\theta_{4}'\rangle=0\rightarrow{\langle\nu_{1}\rangle=\langle\nu_{2}\rangle=0}$ and $\langle\theta_{2}\rangle=0$ we end up with a number of solutions. The most palatable solution gives the following relations between the VEV's,

 \begin{equation}
\langle\theta_{\alpha}\rangle^{2}\equiv{\alpha^{2}}=2\frac{\lambda_{1}\mu_{3}}{\lambda_{2}\mu_{1}}\gamma_1 \gamma_2
\label{alphavev}
 \end{equation}

 \begin{equation}
a^{2}_{1}=\frac{\mu_{1}\mu_{3}}{2\lambda_{1}\lambda_{2}}\frac{\gamma_{1}
\langle\theta_{1}\rangle}
{\gamma_{2}}\quad{and}\quad a^{2}_{2}=\frac{\mu_{1}\mu_{3}}{2\lambda_{1}\lambda_{2}}\frac{\gamma_{2}
\langle\theta_{1}\rangle}
{\gamma_{1}}
\label{avev}
 \end{equation}

 \begin{equation}
\langle\theta_{3}\rangle=\frac{\mu_2}{\mu_3}\langle\theta_{1}\rangle
\label{3vev}
 \end{equation}

 \noindent with all the other singlet VEV's equal to zero, except the $\langle\theta_{\beta}\rangle$ which will be designated by D-flatness condition. Notice that equation (\ref{alphavev}) gives $\alpha^{2}=2\gamma_{1}\gamma_{2}$ for
 $\lambda_{1}\mu_{3}=\lambda_{2}\mu_{1}$. We should also observe that
 combining the equations in (\ref{avev}) we have $a_{1}\gamma_{2}=\pm a_{2}\gamma_{1}$.

 \subsection{D-flatness}
 The D-flatness condition for an anomalous $U(1)$ is given by

\begin{equation}
\sum_{i,j}Q_{ij}^{A}\left(\vert\langle\theta_{ij}\rangle\vert^{2}-\vert\langle\theta_{ji}\rangle\vert^{2}\right)=-\frac{TrQ^{A}}{192\pi^{2}}g_{s}^{2}M_{s}^2
\label{Dc1}
\end{equation}

 \noindent where $Q_{ij}^{A}$ are the singlet charges and the trace $TrQ^{A}$ is over all singlet and non-singlet states. The D-flatness conditions must be checked for each the $U(1)'s.$ In our case we have a $D_{4}$ symmetry and one $U(1)$. The trace in the $SU(5)$ case has the general form

 \begin{equation}
 TrQ^{A}=5\sum{n_{ij}(t_{i}-t_{j})}+10\sum{n_{k}t_{k}}+\sum{m_{ij}(t_{i}-t_{j})}.
\end{equation}

\noindent The coefficients $n_{ij}$, $n_{k}$ and $m_{ij}$ corresponds to the $M_{U(1)}$  multiplicities. Only the curves with a $t_{5}$ charge contributes to the relation since the $t_{l=1,2,3,4}$ are subject to the $D_{4}$ rules. Using this information, the computation of the trace gives:

\begin{equation}
TrQ=(m_{\alpha}'+m_{\beta}'+2m_{\gamma}'-m_{\alpha}-m_{\beta}-2m_{\gamma}-5)t_{5}
\end{equation}

\noindent where the $m_{i},m_{i}'$ are the (unknown)  multiplicities of the singlets $\theta_{i}$ and $\theta_{i}'$, with $i=\alpha,\beta,\gamma$. Inserting the trace in the relation (\ref{Dc1}) we end up with the following equation

\begin{equation}
|\theta_{\alpha}'|^{2}-|\theta_{\alpha}|^{2}+|\theta_{\beta}'|^{2}-|\theta_{\beta}|^{2}+|\theta_{\gamma}'|^{2}-|\theta_{\gamma}|^{2}=(5-\tilde{m}_{\alpha}-\tilde{m}_{\beta}-2\tilde{m}_{\gamma})\mathcal{X}
\end{equation}

 \noindent where $\tilde{m}_{i}\equiv{m_{i}'-m_{i}}$ and $\mathcal{X}=\frac{g_{s}^{2}M_{s}^{2}}{192\pi^{2}}$. By using the results of the F-flatness conditions the  last relation takes the form

 \begin{equation}
 \alpha^{2}+\beta^{2}+2\gamma_{1}\gamma_{2}=(\tilde{m}_{\alpha}+\tilde{m}_{\beta}+2\tilde{m}_{\gamma}-5)\mathcal{X}
 \end{equation}

\noindent which gives an estimation for the $\beta$ VEV ,

\begin{equation}
\beta^{2}=\tilde{\mathcal{M}}\mathcal{X}-\left(1+\frac{\mu_{1}\lambda_2}{\mu_{3}\lambda_{1}}\right)\alpha^{2}\approx{\tilde{\mathcal{M}}\mathcal{X}-2\alpha^{2}}
\label{betavev}
\end{equation}

\noindent where we make use of the equation (\ref{alphavev}) and the approach  $\lambda_{1}\mu_{3}\approx\lambda_{2}\mu_{1}$ in the last step. Finally for shorthand we have set $\tilde{\mathcal{M}}\equiv{\tilde{m}_{\alpha}+\tilde{m}_{\beta}+2\tilde{m}_{\gamma}-5}$. Checking equation (\ref{betavev}) we see that $\tilde{M}$ is a positive number and as a result $\tilde{m}_{\alpha}+\tilde{m}_{\beta}+2\tilde{m}_{\gamma}>5$.

Summarizing, equations (\ref{alphavev},\ref{avev},\ref{3vev}) and  (\ref{betavev}) show us that controlling the scale of $\gamma_{1,2}$ and $\langle\theta_{1}\rangle$  we can have an estimation of the scale of all the singlets participating in the model.

\section{An Alternative Polynomial}
\noindent Another resolvent cubic that shares its discriminant with the quartic polynomial can be built using the following three roots:

\begin{equation}
z_{1}=(t_{1}+t_{2})(t_{3}+t_{4}),\quad{z_{2}=(t_{1}+t_{3})(t_{2}+t_{4})},\quad{z_{3}=(t_{1}+t_{4})(t_{2}+t_{3})}
\end{equation}

\noindent with the two symmetric polynomial set-ups related as follows :
\begin{equation}
z_{1}=x_{2}+x_{3},\quad{z_{2}=x_{1}+x_{3}},\quad{z_{3}=x_{1}+x_{2}}.
\end{equation}
\noindent To see that the two discriminants coincide, note that the differences for each set of symmetric polynomials are related as:
\be x_i-x_j = -(z_i-z_j)\ee
and since the discriminant can be expressed as products of these difference it is trivial to see that the two must coincide:
\be \Delta =\prod_{i\ne j} (z_i-z_j)= \prod_{i\ne j}(x_i-x_j)\,.\ee
In this case the cubic resolvent polynomial has the form:
\begin{equation}
g(s)=a_{5}^{-3/2}[(a_{5}s)^{3}-2a_{3}(a_{5}s)^{2}+(a_{3}^{2}+a_{2}a_{4}
-4a_{1}a_{5})a_{5}s+(a_{2}^{2}a_{5}-a_{2}a_{3}a_{4}+a_{1}a_{4}^{2})]\,.
\label{cubic2}
\end{equation}
And we can see that by setting $g(0)=0$ we obtain the following condition:
\be a_{2}^{2}a_{5}-a_{2}a_{3}a_{4}+a_{1}a_{4}^{2}=0\,.\ee
Substituting  the above condition in the equation of the fives the result is zero, which is not a surprising result since the three symmetric functions of the roots, $z_{i}$,
can be used to rewrite the equation of the GUT fives as:
\begin{equation}
P_{5}=\prod_{i,j}(t_{i}+t_{j})=z_{1}z_{2}z_{3}\prod_{i}^{4}(t_{i}+t_{5})=-g(0)\prod_{i}^{4}(t_{i}+t_{5}).
\end{equation}

If we substitute this new condition into the discriminant we find that it now reads:
\be\Delta \propto 4 \left(4 a_1 a_5-a_2 a_4\right) \left(a_3^2+a_2 a_4-4 a_1 a_5\right){}^2\ee

Combined with the constraint for tracelessness of the GUT group\footnote{$\left\{a_4\to a_0 a_6,a_5\to -a_0 a_7\right\}$},  $b_1=0$, the condition becomes:
\be g(0)=0\to  a_7 a_2^2+ a_3 a_6 a_2=a_0 a_1 a_6^2\,.\ee
Correspondingly the fives of the GUT group now have an equation that factors into only two parentheses,
\be P_5= \left(a_7 a_2^2+a_3 a_6 a_2-a_0 a_1 a_6^2\right) \left(a_3 a_6^2+a_7 \left(a_2 a_6+a_1
   a_7\right)\right)\to P_aP_b\,,  \ee
   where, the first factor vanishes due to the constraint and corresponds to the  roots
  $z_1z_2z_3=0$.\\

  \noindent In this relation it is clear that the trivial condition $g(0)=0$ automatically leads to $P_{5}=0$.  So we need a more general factorisation for the cubic polynomial. In general a cubic is reducible if it can be factorised as a linear and a quadratic part.

\section{Matter Parity From Geometric Symmetry}
\label{D}
One of the major issues in supersymmetric GUT model building is the appearance of  dimension four violating  operators leading to proton decay at unacceptable rates.  The problem is usually solved by introducing the concept of R-parity which is a suitable discrete symmetry preventing the appearance of baryon and lepton four-dimensional non-conserving operators in the Lagrangian.  R-parity is equivalent to a $Z_2$  symmetry, which is the simplest possibility. However, other discrete symmetries in more involved models may be useful as well. The implementation of such a scenario in String and F-theory models in particular has been the subject of considerable recent work \cite{Ibanez:2012wg}-\cite{Honecker:2013hda}.\\

 In our present approach we have constantly dealt with non-Abelian discrete symmetries which were used to organise the fermion mass hierarchies and in particular the neutrino mass textures aiming to reconcile the current experimental data. At the same time, they  are also expected to suppress flavour changing operators. Phenomenological investigations however, have shown that additional discrete symmetries may account for the rare flavour decays in a more elegant manner. This fact could be used as an inspiration to search for  discrete symmetries of different origin in the present constructions.\\

Indeed, a  thorough study of the effective F-theory models the last few years has uncovered a plentiful source of such symmetries which may arise from the internal  geometry and the fluxes.  We will present  such a  mechanism (firstly proposed in~\cite{Hayashi:2009bt} and implemented on specific GUT constructions in~\cite{Antoniadis:2012yk}) in what follows. \\

In constructing models in F-theory the relevant data originate form the geometric properties of the Calabi-Yau four-fold $X$  and the $G_4$-flux. Therefore, if we wish to obtain a $Z_2$ (or some other discrete) symmetry  of geometric origin, in principle we need   to  impose it  on the ($X,G_4$) pair.  It is not easy to prove the existence of such symmetries globally. Nevertheless for the local model constructions  we are interested in it is sufficient to work out such a symmetry in the local geometry around the GUT divisor $S_{GUT}$, which in our case corresponds to an $SU(5)$ singularity. This incorporates the concept of the spectral surface.\\

Indeed, in the weakly coupled  limit of F-theory, the supersymmetric configurations of the effective theory can be described in terms of  the adjoint scalars and the gauge fields. A convenient simplification is based on the  spectral cover description where  the Higgs is replaced by its eigenvalues and the bundle by the corresponding eigenvectors. Since our primary interest is the reduction of $E_8$ to $SU(5)\times SU(5)_{\perp}$ we focus in $SU(5)$ group where the spectral surface is described by the equation:
 \be  \sum_{k=0}^5b_ks^{5-k}=0\,.\ee

 We  consider the GUT divisor $S_{GUT}$ and  three open patches $S,T,U$ covering $S_{GUT}$; we define a phase $\phi_N=\frac{2\pi}{N}$ and a  map $\sigma_N$ such that:
\be  \sigma_N: [S:T:U]\; \to\; [e^{i\phi_N}S: e^{i\phi_N}T:U]\,.\ee
 For a $Z_2$ symmetry discussed in~\cite{Hayashi:2009bt}  one requires a $Z_2$ background configuration, with a $Z_2$ action so that the mapping is:
\be \sigma_2 : [S:T:U]\to [-S:-T:U]\; {\rm or}\; [S:T:-U] \,.\ee
 To see if this is a symmetry of the local geometry for a given divisor, we take local coordinates for the three trivialization patches. These can be defined as $(t_1,u_1) = (T/S,U/S)$, $(s_2,u_2)=(S/T,U/T)$ and $(s_3,t_3)=(S/U,T/U)$. Assuming that $\sigma_2(p)$, is the map of a point  $p$ under $\sigma_2$ transformation, the corresponding local coordinates are mapped according to
\be \begin{split}
\left.(t_1,u_1,\xi_s)\right|_{\sigma_2(p)}&=  (t_1,-u_1, -\xi_s)|_{p} \\
\left.(s_2,u_2,\xi_t)\right|_{\sigma_2(p)}&=  (s_2,-u_2, -\xi_t)|_{p}\\
\left.(s_3,t_3,\xi_u)\right|_{\sigma_2(p)}&=  (-s_3,-t_3,\xi_u)|_{p}
\end{split}\ee

This is an $SU(3)$ rotation on the three complex coordinates which  acts on the spinors in the same way. Hence, starting from a  $Z_2$ symmetry of the three-fold we conclude that a $Z_2$ transformation is also induced on the spinors. The required discrete symmetry must be a symmetry of the local geometry. This can happen  if  the defining equation of the spectral surface is left invariant under the corresponding discrete transformation.  Consequently we expect non-trivial constraints on the polynomial coefficients $b_k$   which carry the information of local geometry.\\

 In order to extract these  constraints we focus on a single trivialization patch  and take $s$ to be the coordinate along the fiber.  Under the mapping of points $p\to \sigma(p)$ we  consider the phase transformation
\be s(\sigma(p)) = s(p)\,e^{i\phi},\; \; b_k(\sigma(p)) \,=\,b_k(p)\,e^{i(\chi-(6-k)\phi)}\,.\ee
Under this action, each term in the spectral cover equation transforms the same way,
\be b_k s^{5-k}\ra e^{i(\chi-\phi)}b_k s^{5-k}\,.\ee
We observe that the spectral surface equation admits a continuous symmetry. A trivial solution arises for $\phi=0$ where all $b_k$ pick up a common phase:
\be s\ra s,\; b_k\ra b_k \,e^{i\chi}\ee
In the general case however, the non-trivial solution accommodates a $Z_N$ symmetry for
\be\phi=\frac{2\pi}{N}.\ee
Thus, for $N=2$, we have $\phi=\pi$  and the trasformation reduces to
\ba
s\ra -s,\; b_k\ra (-1)^{k} e^{i\chi}\,b_k\quad{.}\label{bphase}
\ea
\subsection{Extension to $C_5\rightarrow C_4\times C_1$}
In the event that the spectral cover is taken to split down to products of factors, for example $C_5\rightarrow C_4\times C_1$, this symmetry is conveyed to the matter curves by consistency with the original spectral cover equation. It is trivial to determine that the coefficients of $C_5$ are related to the $C_4\times C_1$ coefficients by:
\be b_k=\sum_{n+m=12-k}a_ia_j
\ee
where $i\neq j$. As such, we can  directly write that if
\be a_{n} \to  \text{e}^{i\psi_{n}} \text{e}^{i (3-n)\phi} a_{n} \ee
so that the product  $a_{n} a_{m}$ picks up a total phase:
\be a_{n} a_{m} \to  \text{e}^{i(\psi_{n}+\psi_{m})} \text{e}^{i (6-n-m)\phi} a_{n}a_m= \text{e}^{i(\psi_{n}+\psi_{m})} \text{e}^{-i (6-k)\phi} a_{n}a_m\ee
then provided the phases of the $a_n$ coefficients satisfy $\chi= \psi_{n}+\psi_{m}$, the symmetry is handed down to the split spectral cover. This is trival to enforce since the phases  are independent of the index $k$. It can also be demonstrated that this consistency requires the coefficients of $C_4\times C_1$ to have phases in two cycles: $\psi_i=\psi_1=\psi_2=\dots=\psi_5$ and $\psi_j=\psi_6=\psi_7$, in order to be consistent with the $C_5$ phase.

\tref{exparity} shows some examples of possible parities we might assign to the $C_4\times C_1$ coefficients. In most cases, the minimal $N=2$ scenario will be the most appealing and manageable choice, though this mechanism is not confined to it.

\begin{table}[t!]\centering
\begin{tabular}{|c|c|c|c|c|}\hline
$a_n$&$N=2$&$N=3$&$N=4$&$N=5$\\\hline
$a_1$&$-$&$\alpha^2$&$\beta^2$&$\gamma^2$\\
$a_2$&$+$&$\alpha$&$\beta$&$\gamma$\\
$a_3$&$-$&$1$&$1$&$1$\\
$a_4$&$+$&$\alpha^2$&$\beta^3$&$\gamma^4$\\
$a_5$&$-$&$\alpha$&$\beta^2$&$\gamma^3$\\
$a_6$&$+$&$1$&$\beta$&$\gamma^2$\\
$a_7$&$-$&$\alpha^2$&$1$&$\gamma$\\\hline
\end{tabular}
\caption{$Z_N$ parities coming from geometric symmetry of the spectral cover. In the case of $C_5\to C_4\times C_1$, a general phase relates the parities of $a_{1,2,3,4,5}$, such that if we flip the parity of $a_1$ all the other $a_i$ in this chain must also change. A similar rule applies to $a_{6,7}$.\label{exparity}}\end{table}

   \end{document}